%% file: main.tex
\documentclass[a4paper,11pt]{article}
\usepackage{jinstpub}
\usepackage{lineno}
\usepackage{siunitx}
\usepackage{subcaption}
\usepackage{tikz}
\usepackage{xspace}

\title{\boldmath Energy, time, and position resolution measurements of an array of large tapered LYSO crystals}

\author[a]{O.~Beesley}
\author[b]{J.~Carlton}
\author[c]{D.~Ding}
\author[d]{S.~Foster}
\author[e]{K.~Frahm}
\author[e]{T.~Frank}
\author[f]{L.~Gibbons}
\author[e]{D.~Goeldi}
\author[b]{T.~Gorringe}
\author[a]{D.W.~Hertzog}
\author[e,1]{S.~Hochrein}\note{Corresponding author.}
\author[g]{J.~Hui}
\author[i]{E.~Klemets}
\author[a]{J.~LaBounty}
\author[g,h]{J.~Liu}
\author[j]{P.~Lopez Maggi}
\author[f]{W.~Osar}
\author[a]{R.~Roehnelt}
\author[a]{P.~Schwendimann}
\author[g]{W.~Shuai}
\author[e]{A.~Soter}
\author[a]{E.~Swanson}

\affiliation[a]{Center for Nuclear Physics and Astrophysics (CENPA), University of Washington \\Seattle, USA}
\affiliation[b]{University of Kentucky \\Lexington, USA}
\affiliation[c]{Shanghai Institute of Ceramics, Chinese Academy of Sciences (SICCAS) \\Shanghai, China}
\affiliation[d]{Amherst College \\Amherst, USA}
\affiliation[e]{ETH Zurich \\Zurich, Switzerland}
\affiliation[f]{Cornell University \\Ithaca, USA}
\affiliation[g]{Shanghai Jiao Tong University \\Shanghai, China}
\affiliation[h]{New Cornerstone Science Laboratory, Tsung-Dao Lee Institute, Shanghai Jiao Tong University \\Shanghai, China}
\affiliation[i]{University of British Columbia \\Vancouver, Canada}
\affiliation[j]{Department of Physics, FCEN, University of Buenos Aires \\Buenos Aires, Argentina}

\emailAdd{shochrein@ethz.ch}

\abstract{We report on the performance of six custom-made tapered LYSO crystals of unprecedented volume, which constitute a sector of the 19 radiation length electromagnetic calorimeter planned for the PIONEER experiment. The longitudinal response uniformity of each crystal was measured using radioactive sources before characterizing the energy and time resolution of the crystals in an array using a 20 to \SI{80}{MeV} positron beam at the Paul Scherrer Institute. The array demonstrated an energy resolution better than \SI{2}{\percent} for energies above \SI{40}{MeV} and a time resolution better than \SI{140}{ps} at \SI{30}{MeV}. The spatial resolution was measured in the central region of the array to be \SI{5.06}{mm} in x and \SI{5.20}{mm} in y at 70 MeV, after removal of local biases. The measured properties satisfy the key design parameters of the PIONEER calorimeter for the measurement of rare pion decays.}


\begin{document}

\newcommand{\Gfour}{\textsc{Geant}4\xspace}

\maketitle
\flushbottom

\include{Sections/introduction}

\include{Sections/the_pioneer_electromagnetic_calorimeter}

\include{Sections/six_crystal_prototype}

\include{Sections/lyso_array_beam_test}

\include{Sections/results}

\include{Sections/conclusion}

\appendix
\include{Sections/app-pmt}
\include{Sections/app-calibration}

\acknowledgments

 This work was supported by the United States Department of Energy grants DE-FG02-97ER41020 and DE-SC0008037 as well as the Swiss National Science Foundation (SNSF) project grant No. 200021\_215301. We thank the PSI accelerator staff for their support and delivery of a reliable beam at multiple energies. SICCAS has been instrumental in the research and development of the custom LYSO crystals used in our testing. We also thank our PIONEER colleagues for their support and feedback.


\bibliographystyle{JHEP}
\bibliography{biblio.bib}

\end{document}

%% file: Sections/introduction.tex
\section{Introduction}
\label{sec:intro}

The Phase\;I goal of the PIONEER experiment~\cite{PIONEER:2022yag} at the Paul Scherrer Institute will test lepton flavor universality to one part in $10^4$ by determining the ratio $R_{\pi} = \Gamma(\pi\to e\nu)/\Gamma(\pi\to\mu\nu)$. The rare $\pi \rightarrow e $ decay channel ($R_{\pi} \sim 1.25\times 10^{-4}$) results in the emission of a \SI{69.8}{MeV} monoenergetic positron, while a Michel positron in the range from 0 to \SI{52.8}{MeV} arises from the dominant muon decay channel. While these final states should be distinct in energy, the challenges lie in detector resolution and containment effects in this relatively low-energy regime. An energy resolution better than \SI{2.5}{\percent} at \SI{50}{MeV} is required to achieve sufficient separation of the decay channels in the calorimeter. To obtain the required statistics, a positive pion stop rate greater than \SI{300}{kHz} must be used in Phase I. It is anticipated that a rate greater than \SI{3}{MHz} will be required during a later phase to measure the $10^{-8}$ pion beta decay branch that determines $V_{ud}$.  These rates result in a high occupancy of the electromagnetic calorimeter that is used to record the times and energies of positrons from pion and muon decays. This motivates calorimeter segmentation in conjunction with a time resolution better than \SI{200}{ps} to mitigate pileup effects.

%% file: Sections/the_pioneer_electromagnetic_calorimeter.tex
\section{The PIONEER electromagnetic calorimeter}
\label{sec:intro_calorimeter}

\begin{figure}[htbp]
\vspace{0.5cm}
\centering
\begin{subfigure}{0.4\textwidth}
    \centering
    \includegraphics[width=\linewidth]{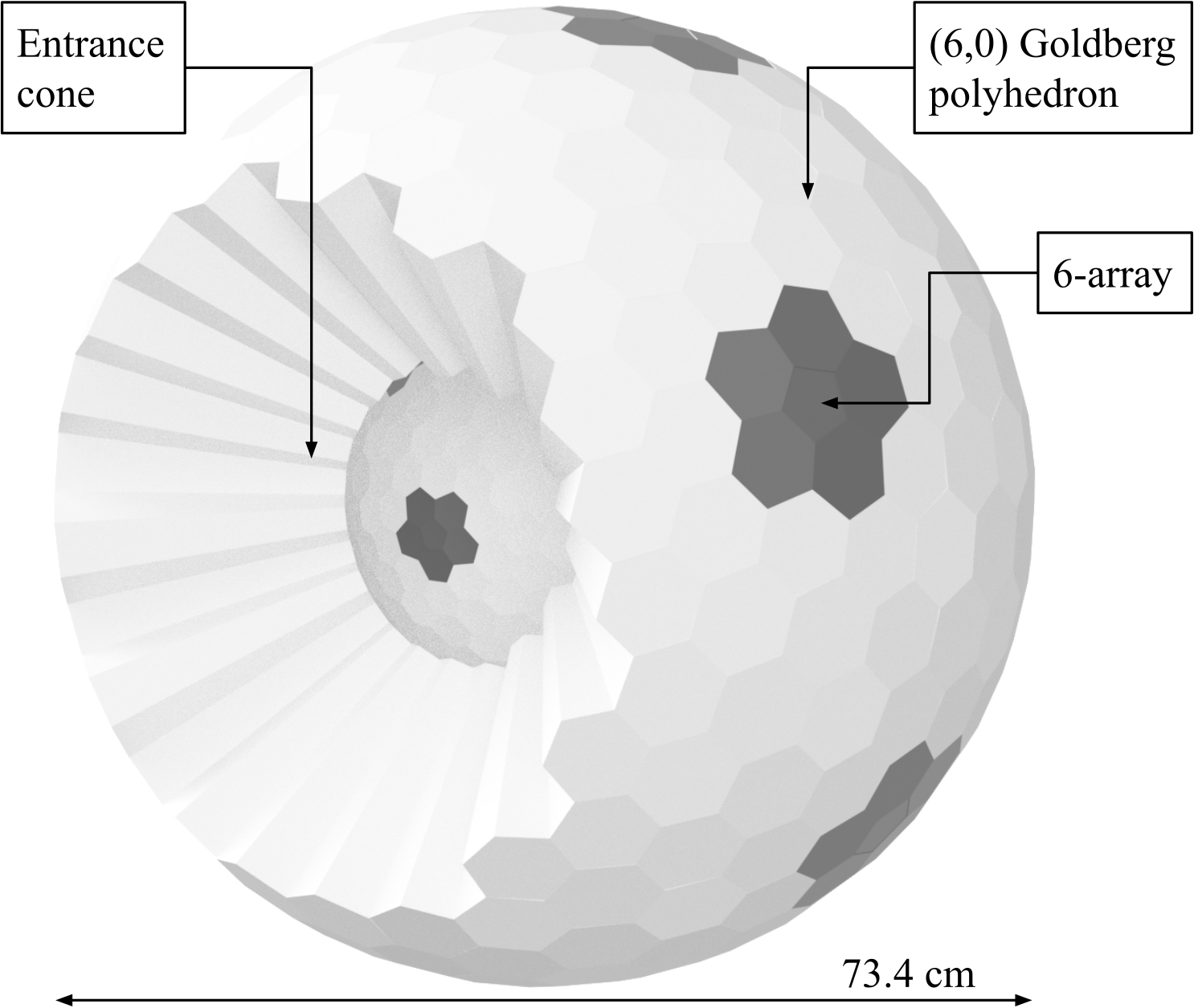}
    \caption{}
    \label{fig:Setupa}
\end{subfigure}
\qquad
\begin{subfigure}{0.4\textwidth}
    \centering
    \includegraphics[width=\linewidth]{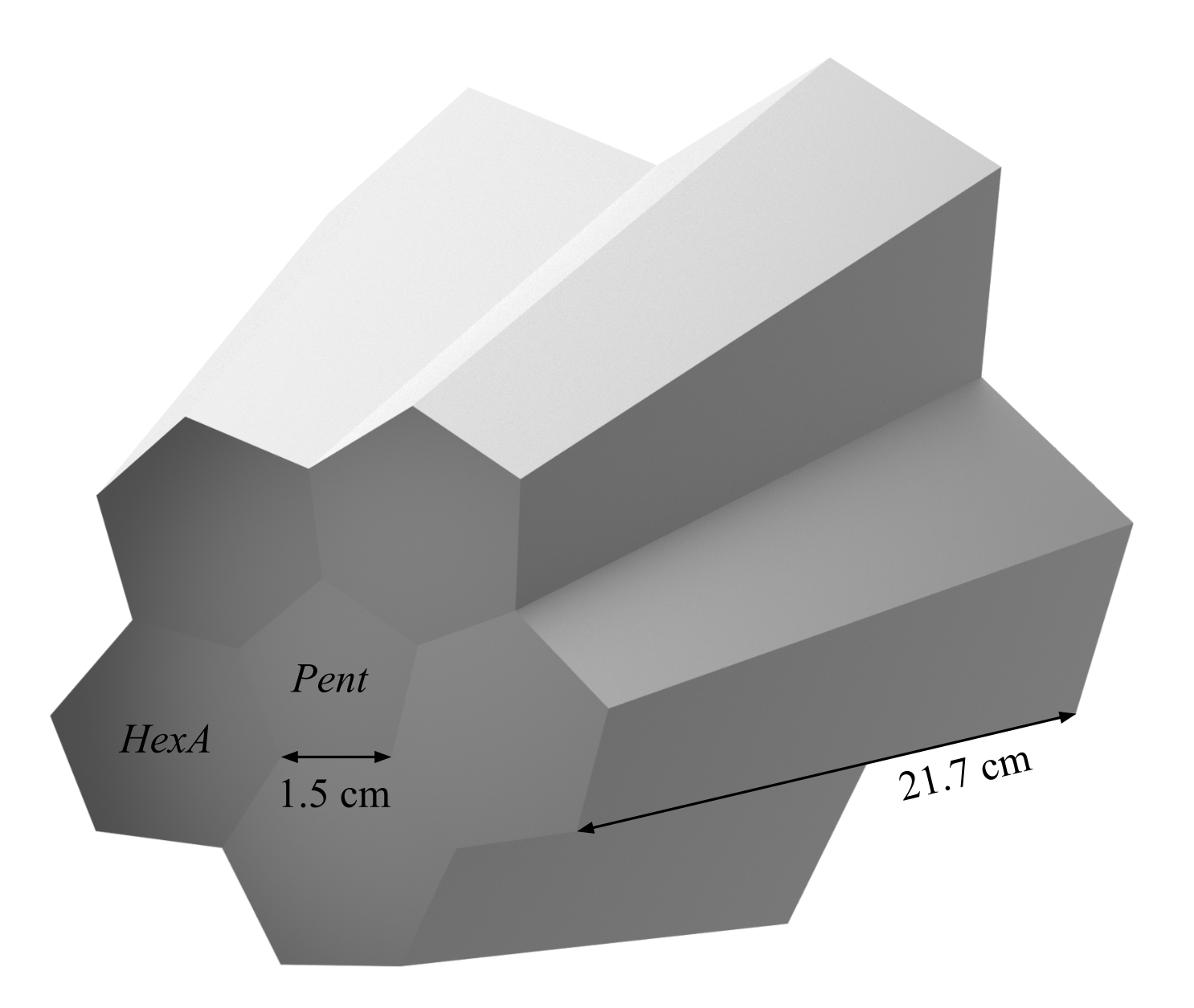}
    \caption{}
    \label{fig:Setupb}
\end{subfigure}

\vspace{0.2cm}

\begin{subfigure}{0.4\textwidth}
    \centering
    \includegraphics[width=\linewidth]{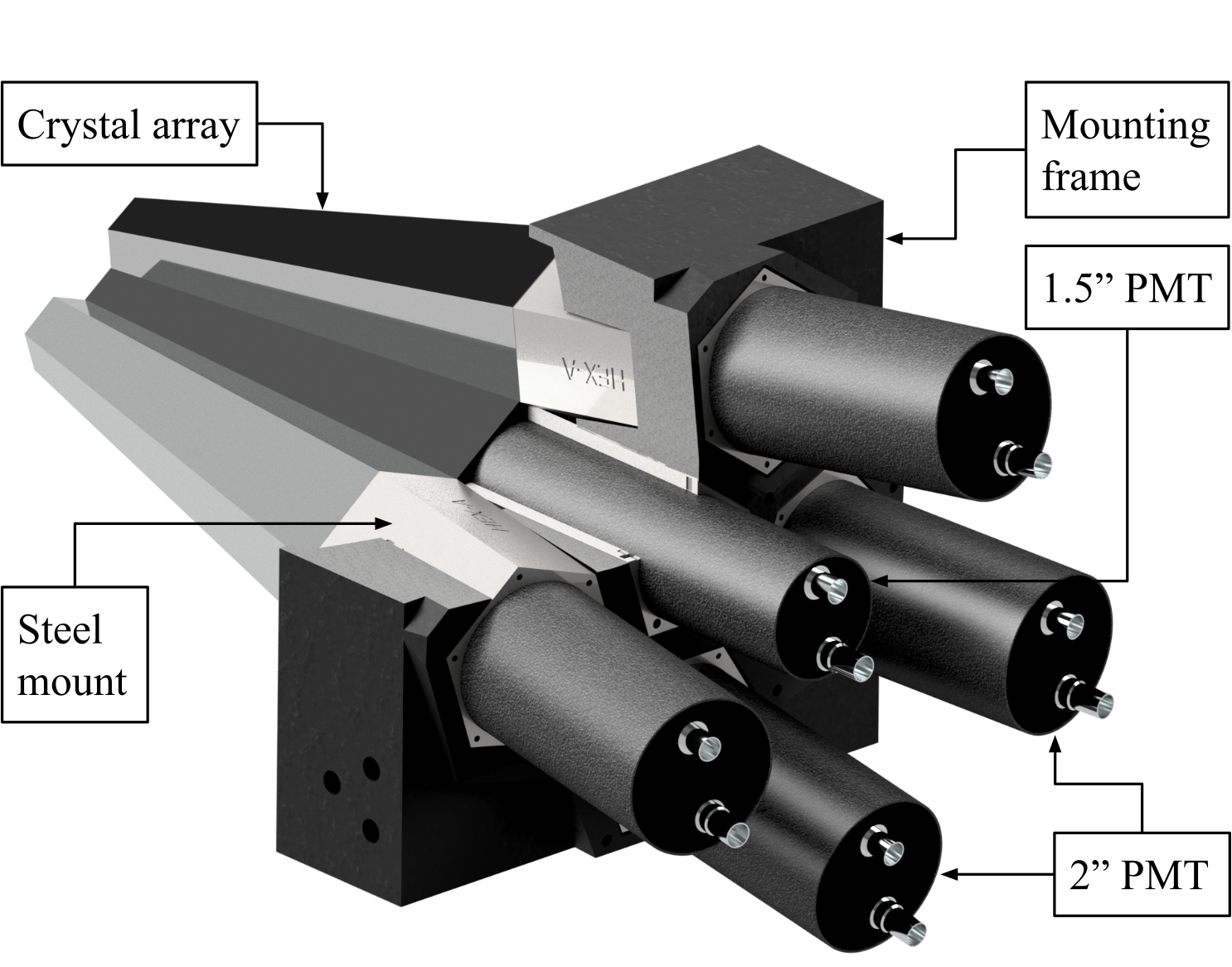}
    \caption{}
    \label{fig:Setupc}
\end{subfigure}
\qquad
\begin{subfigure}{0.4\textwidth}
    \centering
    \includegraphics[width=\linewidth]{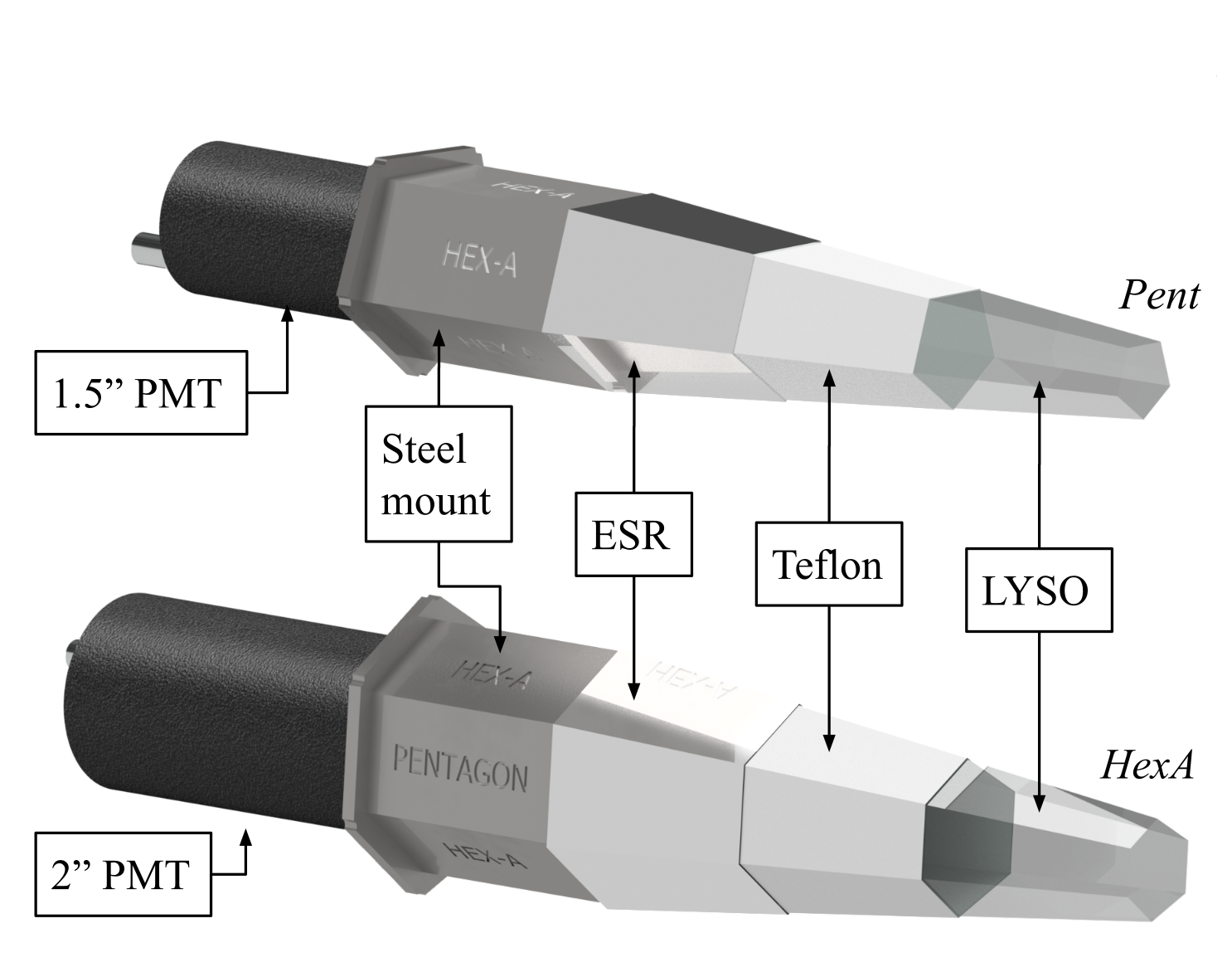}
    \caption{}
    \label{fig:Setupd}
\end{subfigure}

\caption{The top row shows (a) a 3D model of the initial PIONEER calorimeter design based on the (6,0) Goldberg polyhedron, alongside (b) the six crystal sub-cluster assembled for this work. The bottom row shows (c) a section view of the mechanical assembly and (d) a breakdown of the wrapping layers of the LYSO crystals as used during the beam test.}
\label{fig:Setup}
\end{figure}

Cerium doped lutetium-yttrium oxyorthosilicate (LYSO) is a calorimeter material with high light yield (\SI{32000}{photons/MeV}\footnote{As measured by SICCAS at the production level.}), relatively fast decay time (\SI{40}{ns}), high density (\SI{7.4}{g/cm^3}), and high stopping power (radiation length $X_0 = $ \SI{1.14}{cm} and Molière radius $R_M = $ \SI{2.07}{cm}) \cite{Mao20082425}. In 2023, we tested the performance of an array of ten \qtyproduct[product-units = power]{2.5 x 2.5 x 18}{cm^3} rectilinear LYSO crystals. The measurements demonstrated an energy resolution of \SI{1.52}{\%} at \SI{70}{MeV} and a time resolution of \SI{110}{ps} at \SI{30}{MeV}, validating LYSO as a calorimeter material for PIONEER \cite{Beesley:2025lyso}. However, the full calorimeter requires a spherical geometry that cannot be composed of rectilinear shapes. In the initial LYSO crystal calorimeter design that drove this study, 311 crystals form a truncated icosahedral polyhedron geometry with an upstream opening when assembled together. This geometry, known as the (6,0) Goldberg polyhedron, is composed of one extruded pentagonal shape denoted $\textit{Pent}$ and seven extruded hexagonal shapes denoted $\textit{HexA-HexG}$ ordered by size, where the $\textit{HexA}$ crystal is the smallest (besides the $\textit{Pent}$) and the $\textit{HexG}$ crystal is the largest. A model of the geometry is displayed in figure \ref{fig:Setupa}. Each of these crystal shapes is larger by volume\footnote{The \textit{Pent} crystal volume is approximately \SI{265}{cm^3} and is the smallest of the shapes.} than any LYSO crystal that has been previously produced and tested \cite{Mao:2012dr, Bartocci:space_LYSO}. The PIONEER calorimeter design is currently transitioning to a modified (5,1) Goldberg polyhedron geometry, optimized for more uniform crystal shapes, allowing all crystals to be read out by the same PMT.

Tapered crystal shapes introduce complications in production, assembly, and performance that are not present in the rectilinear shapes previously tested. Large LYSO crystals are grown using the Czochralski method \cite{Cooke:2000lyso, Yoshikawa:2013czochralski}, but due to the relatively high melting point of LYSO (\SI{2100}{\degreeCelsius}) and its brittleness, the growth of high-quality boules sufficiently large to maintain the shapes required for the PIONEER calorimeter is a significant production challenge. These large crystals must then be assembled into a crystal ball segment with minimal gaps and no dead supporting material on the front faces to avoid ionizing energy loss from positrons entering the crystal. These engineering constraints present an assembly challenge that necessitates rear-mounting of the crystals with minimal coverage of the back face to allow for light readout. Additionally, the tapered geometry causes scintillation photons traveling from the narrow end toward the wide readout face to impinge on the tapering side walls at increasingly grazing angles. The light reaching the back of the crystal thereby has a smaller angle of incidence with the readout face and a reduced probability of total reflection; it is effectively focused. As the interaction point moves closer to the photosensor, the produced light undergoes fewer reflections before reaching the readout face and thus the light collection decreases, introducing a longitudinal non-uniformity. This focusing effect can be mitigated by crystal surface treatment and the choice of crystal wrapping \cite{Diehl:2017panda}.

In this work, we describe our characterization of a LYSO calorimeter prototype, an array of six large tapered LYSO crystals, shown in figure \ref{fig:Setupb}. The crystals were grown at the Shanghai Institute of Ceramics, Chinese Academy of Sciences (SICCAS) and tested at the $\pi$M1 beam line of the High Intensity Proton Accelerator Facility (HIPA) at the Paul Scherrer Institute (PSI) using positrons ranging in energy from 20 to \SI{80}{MeV}.

At these energies, the electromagnetic showers occupy a smaller volume and contain substantially fewer particles than those typically considered in electromagnetic calorimetry. The cascades produce only a small number of particles above the critical energy in LYSO of $E_c\sim $ \SI{9}{MeV}, calculated using an effective atomic number of $Z_\text{eff}=66$ \cite{Campana:2020oyk}. Bremsstrahlung and pair production, the processes that drive shower formation, only dominate above this energy, while below energy loss by ionization, excitation, Compton scattering and the photoelectric effect become increasingly significant \cite{Fabjan:2003aq, PDG_2024}. The small number of particles therefore involved results in intrinsically asymmetric shower development, with individual events exhibiting large event-to-event statistical fluctuations in the number, energies, and directions of their secondary particles.

%% file: Sections/six_crystal_prototype.tex
\section{Six crystal prototype}
\label{sec:prototype}

\subsection{Calorimeter prototype}

For the LYSO calorimeter prototype, a sub-cluster consisting of six tapered crystals, one $\textit{Pent}$ surrounded by five $\textit{HexA}$, was chosen. The group represents the smallest symmetric unit of the proposed full geometry. The campaign successfully validated the mechanical assembly concept: the six-crystal prototype was rear-mounted in a prototype of the final honeycomb support structure, via a steel mount glued to the back surface. The hollow steel mount allows the insertion of a \SI{2}{"} PMT for readout of the $\textit{HexA}$ crystals, and a \SI{1.5}{"} PMT for the smaller $\textit{Pent}$ crystal. Figure \ref{fig:Setupc} shows a rendering of the mechanical setup.

Before assembly, the crystal dimensions were measured to high precision (< \SI{1}{\micro m}) using a Mitutoyo KN8 coordinate measuring machine. The crystal surfaces were found to be extremely flat (variance < \SI{2}{\micro m}), while the crystals were designed to be slightly oversized relative to their nominal sizes in the PIONEER calorimeter by \SI{0.2}{mm} to allow for a later retreatment of the crystal surfaces for the final experiment.

The LYSO crystals were read out by one Hamamatsu H3178-51 and five H13719\footnote{Hamamatsu R13089 PMT in a custom made assembly with magnetic shielding and tapered voltage divider circuit.} PMT assemblies, with tapered voltage dividers to increase their dynamic range. The data sheets provided by Hamamatsu claim a linear response (< \SI{2}{\percent} deviation) up to a peak anode current of \SI{150}{mA} and \SI{100}{mA} respectively. For the custom made H13719 PMT assembly, this value was confirmed in a bench test as explained in the end of appendix \ref{app:pmt_gain}. Extrapolating from the peak anode current measured at the \SI{1.33}{MeV} $^{60}$Co line at the nominal \SI{1.5}{kV} operation voltage, it was found that this corresponds to the linear range ending at the light-equivalent of $\sim$\SI{45}{MeV}. Because energy depositions in a single crystal can exceed this value, the PMTs were operated at a reduced voltage during the beam test, as described in section \ref{sec:calibrations}.

\subsection{Uniformity scans}
\begin{figure}[h]
    \centering
\includegraphics[width=0.8\textwidth]{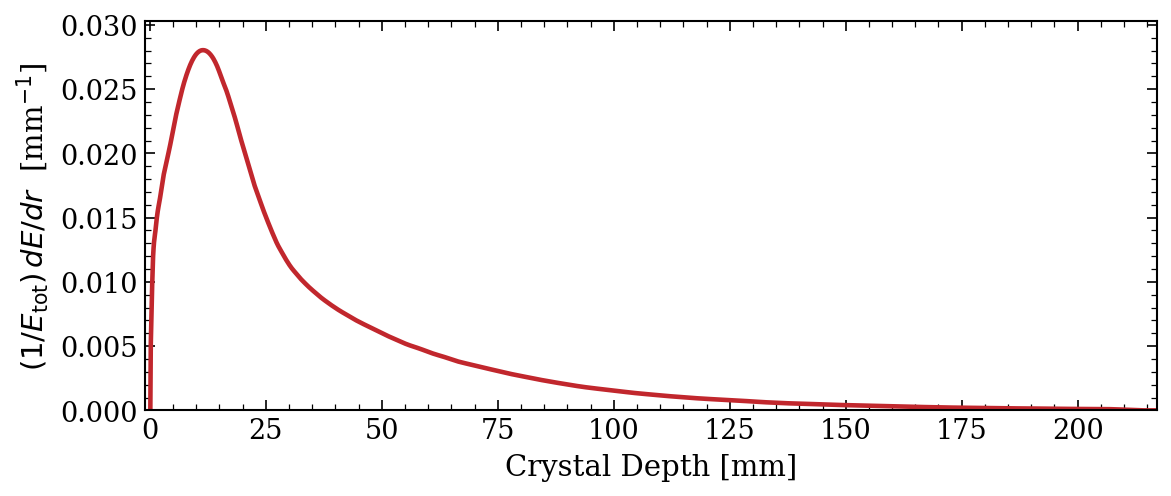}
    \caption{Simulated averaged longitudinal energy-deposition profile of 70 MeV positron showers in the LYSO array, expressed as the fractional energy deposited per unit depth, $(1/E_{tot})\,dE/dr$. The deposition peaks at approximately one radiation length, with a tail at larger crystal depths due to stochastic shower fluctuations.}
    \label{fig:shower_profile}
\end{figure}

\begin{figure}[h]
    \centering
\includegraphics[width=0.8\textwidth]{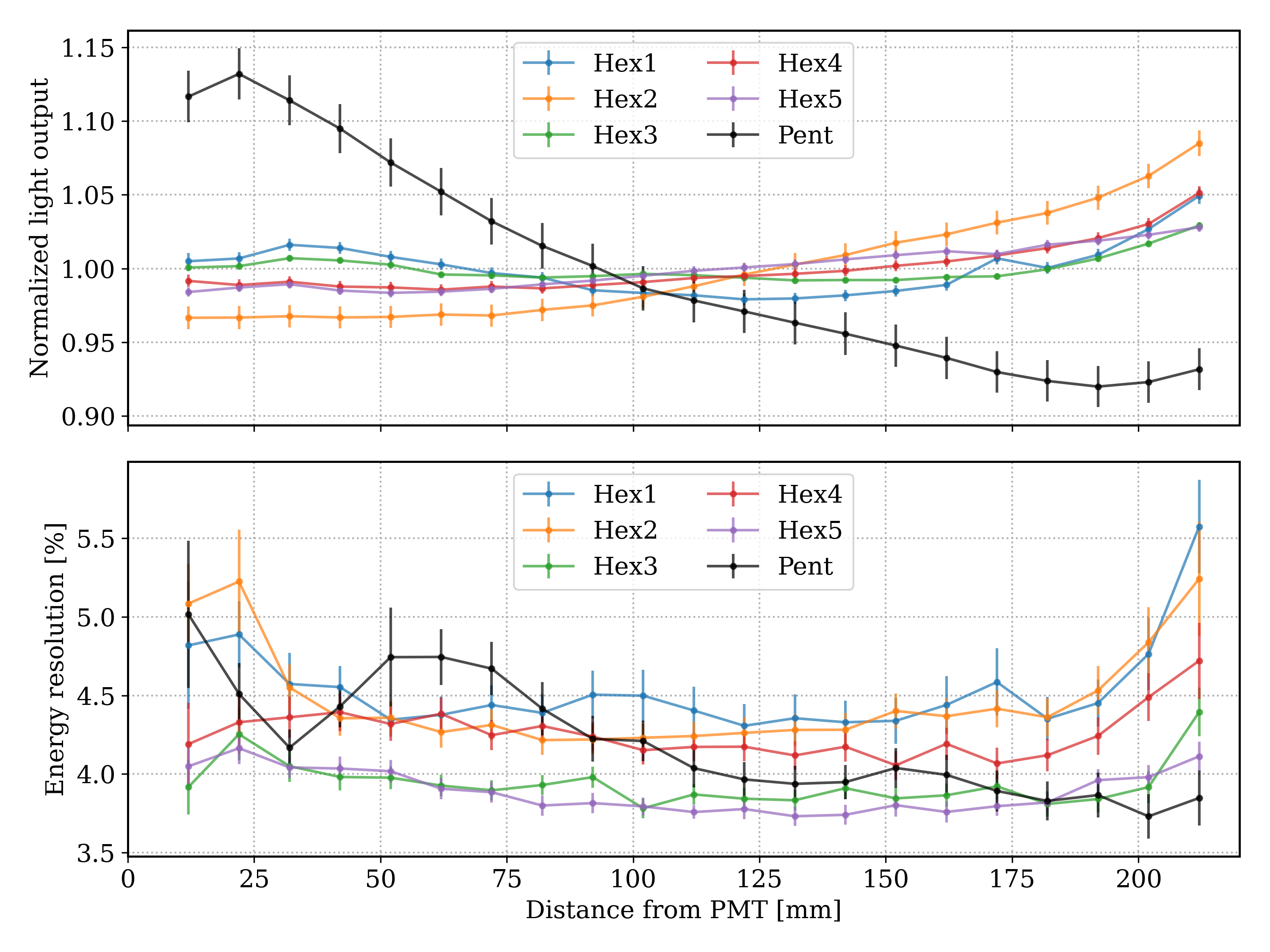}
    \caption{Uniformity scan along the crystal length with a collimated $^{60}$Co source. The data points represent the averages of the two $^{60}$Co photopeak measurements at \SI{1.17}{MeV} and \SI{1.33}{MeV}. The top plot shows for each crystal the collected light along the scan, normalized by their average. While the light output of the \textit{HexA} crystals slightly decreases toward the readout face, the \textit{Pent} crystal exhibits the opposite trend. The longitudinal non-uniformity NUF (standard deviation of the measured light outputs that have been normalized to their mean) is below \SI{7}{\percent} for all crystals. The bottom plot shows the energy resolution which was found to be below \SI{4.5}{\percent} in the center of all crystals, with a slight degradation observed towards the crystal ends.}
    \label{fig:Uniformity}
\end{figure}

Positrons incident on the PIONEER LYSO calorimeter form electromagnetic showers whose energies are deposited over an extended longitudinal profile, as shown in figure \ref{fig:shower_profile} for \SI{70}{MeV} positrons using \Gfour simulations~\cite{Agostinelli:2002geant4, Allison:2006geant4}. Because the crystal's effective light response varies with depth, the measured signal is a convolution of the longitudinal response uniformity and this shower profile. We quantify the longitudinal response non-uniformity (NUF) as the standard deviation of the light outputs along the crystal length that have been normalized to their mean. By folding the shower profile with varied levels of response non-uniformity, we can determine that a NUF > \SI{10}{\percent} would significantly degrade the calorimeter energy resolution in the target energy region, despite the localized energy deposition in the front of the crystals. Therefore, prior to the beam test, each crystal was characterized in the laboratory. To mitigate the focusing effect, the manufacturer roughened one lateral surface of each crystal with a roughness of Ra = $(0.50 \pm 0.05)$\,\si{\micro m}. The crystals were coupled to the PMT using the same configuration as in the beam test. A $^{60}$Co source was collimated with three half-value layers towards the LYSO crystal and mounted on a motorized linear stage, which was used to scan the response along the crystal length. A coincidence trigger with a NaI(Tl) detector placed opposite the source was required to suppress the intrinsic $^{176}$Lu background in LYSO. Several wrapping configurations were tested; the best performance was obtained with a Teflon layer directly around the crystal to diffusely reflect light back into the crystal, covered by 3M Enhanced Specular Reflector (ESR) foil to optically separate the crystals. A visualization of the wrapping layers is shown in figure \ref{fig:Setupd}. As shown in figure \ref{fig:Uniformity}, this procedure resulted in a NUF < \SI{7}{\percent} for all crystals.\footnote{Tests to further improve the longitudinal uniformity are in progress, involving both varying the number of crystal faces that are roughened and the method of wrapping.} Additionally, an average energy resolution below \SI{5}{\percent} was measured in all crystals at the \SI{1.17}{MeV} and \SI{1.33}{MeV} $^{60}$Co photopeaks, with a slight degradation observed towards the crystal ends.

%% file: Sections/lyso_array_beam_test.tex
\section{LYSO array beam test}
\label{sec:beam_test}

\subsection{Beamline}

The LYSO array was characterized at PSI’s $\pi$M1 beam line during one week of beam time in August and September 2025. This secondary beam line is generated by the \SI{50.6}{MHz}, \SI{590}{MeV} proton beam impinging on a \SI{5}{mm} rotating carbon target. Although the target interaction primarily yields pions, the beam line was tuned to select positrons to match the PIONEER signal, produced via neutral pion decay and subsequent pair production. The \SI{21}{m} magnetic channel employs two \SI{75}{\degree} dipoles to create a dispersive intermediate focus; collimation at this stage ensures the precise momentum selection required for intrinsic energy resolution measurements \cite{Cline:2022pim1}. Below \SI{80}{MeV}, the beam composition is dominated by positrons. The remaining background species were rejected in offline analysis using Time-of-Flight (ToF) information relative to the cyclotron RF, exploiting the velocity differences of particles with the same momentum.

\subsection{Beam test setup and beam profile}

\begin{figure}[h]
    \centering
    \includegraphics[width=\textwidth]{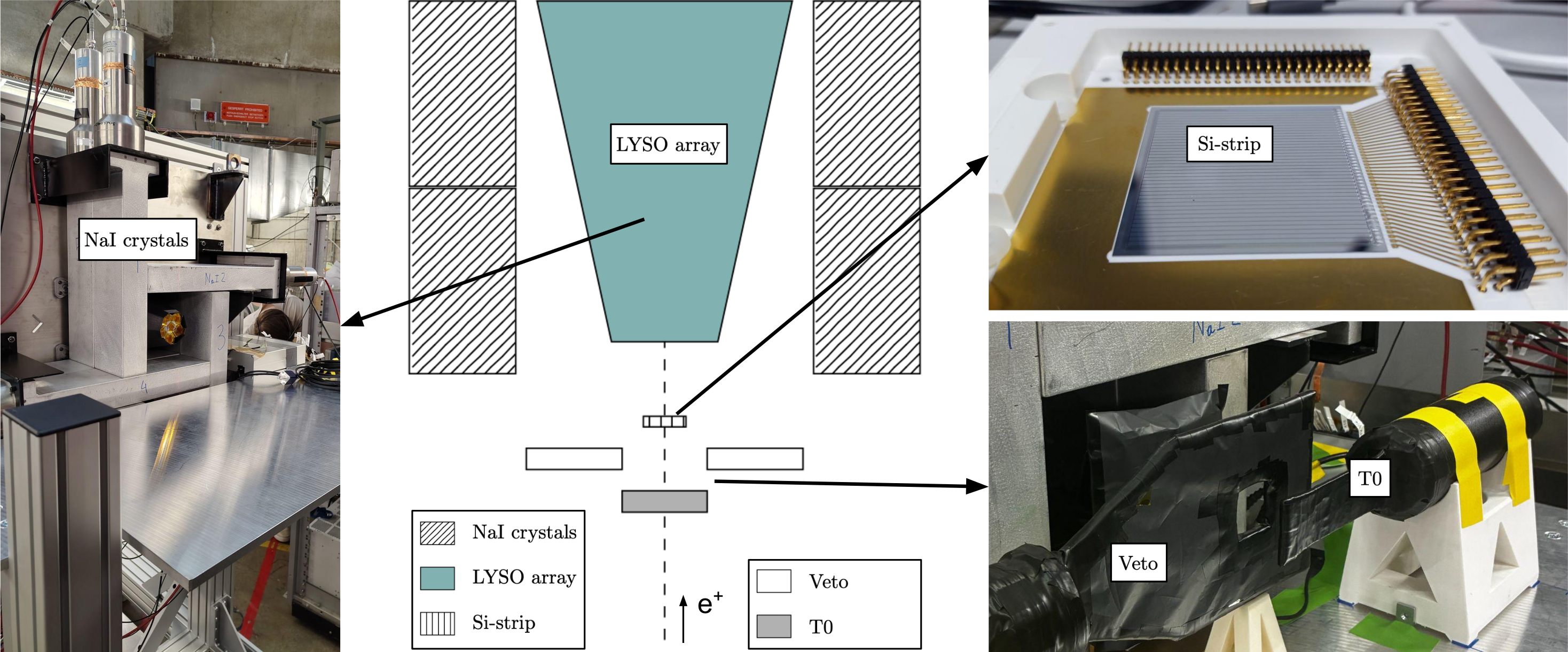}
    \caption{Top view of the setup during the LYSO beam test at PSI with pictures showing the individual detector components.}
    \label{fig:Beamtime_setup}
\end{figure}

The LYSO calorimeter was mounted on an aluminum frame with the \textit{Pent} center axis aligned with the beam axis and the smaller crystal faces pointing in the upstream direction. The tapered LYSO array was surrounded by eight large NaI(Tl) detectors, which were used to catch shower leakage in the lateral directions. The upstream configuration relied on a fixed trigger and tracking system aligned with the beam axis and the center of the \textit{Pent} crystal. The event timing reference was provided by a \qtyproduct[product-units = power]{40 x 40 x 2}{mm^3} T0 scintillator. This was positioned along the same central axis, in alignment with a \qtyproduct[product-units = power]{140 x 140 x 5}{mm^3} veto scintillator with a central \qtyproduct[product-units = power]{40 x 40}{mm^2} aperture. The trigger logic required a coincidence $\mathrm{T0}\, \cdot\,\overline{\mathrm{VETO}}$, selecting only particles that passed through the central acceptance hole without significant scattering in the T0. Downstream of the trigger, a \qtyproduct[product-units = power]{40 x 40 x 1}{mm^3} double-sided silicon strip detector with \SI{1}{mm} pitch was used as a beam hodoscope. On each side, the central 20 strips were read out to provide horizontal and vertical position information. The full setup and its components are shown in figure \ref{fig:Beamtime_setup}.

The tracking system indicated a circular beam spot width of approximately \SI{10.3}{mm} ($\sigma$) for \SI{70}{MeV} positrons passing the Veto. The beam rate varied from 0.3 to \SI{7}{kHz} with increasing beam momentum, and the trigger was throttled to a maximum of $\sim$1.5~kHz via an analog gate to ensure data-taking stability. The beam momentum bite, $\Delta p/p$, was previously determined to be \SI{< 0.65}{\percent} at \SI{70}{MeV/c} for an identical momentum slit width \cite{Beesley:2025lyso}. This upper limit was derived from the arrival time dispersion of muons and pions. A more rigorous measurement found a momentum bite of \SI{0.1}{\percent} at higher beam energies \cite{Cline:2022pim1}. Consequently, momentum spread was assumed to be a subdominant contribution to the final energy resolution.

\subsection{Waveform processing}

\begin{figure}[h]
    \centering
    \includegraphics[width=0.8\textwidth]{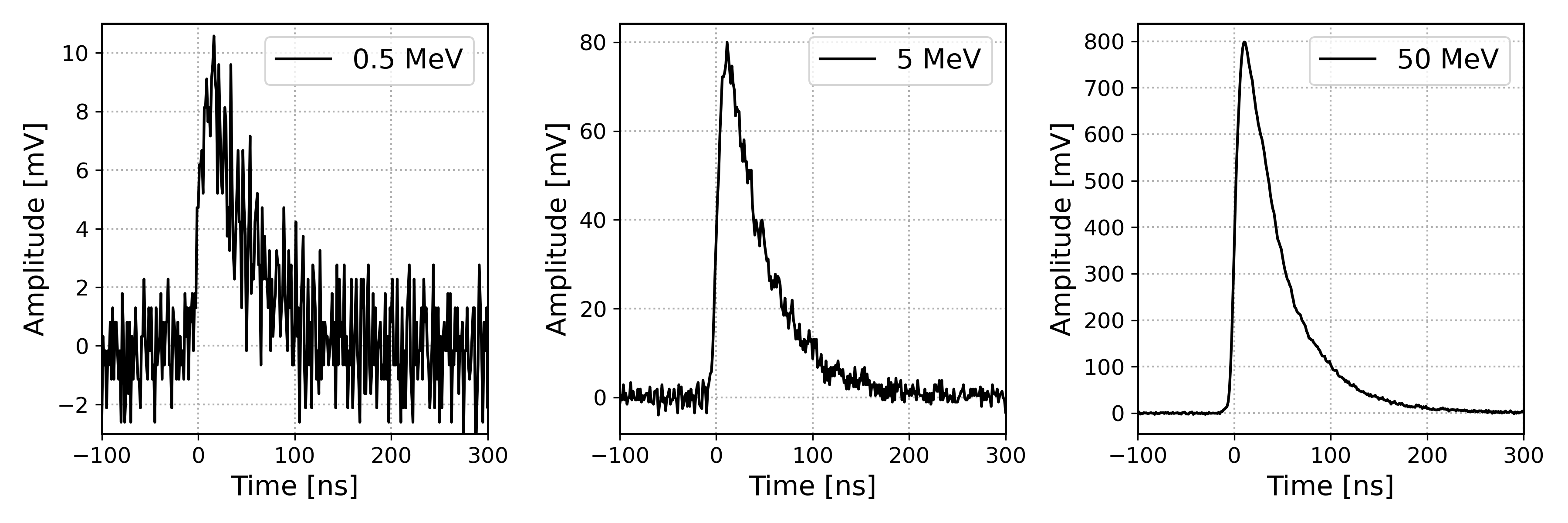}
    \caption{Example waveforms from the \textit{Pent} crystal at different energies. Consistent pulse shapes were observed across a large dynamic range, including pulses corresponding to sub-MeV energy deposits.}
    \label{fig:Waveforms}
\end{figure}

Signal waveforms were recorded by a $\mu$TCA-based digitizer operating at \SI{800}{MSPS} with 12-bit resolution and a \SI{2}{V} dynamic range \cite{Sweigart:2016jty}. A representative sample of waveforms corresponding to varying energy depositions in the central LYSO crystal is presented in figure \ref{fig:Waveforms}. The signal charge was derived by integrating the digitized waveform relative to a pedestal-subtracted baseline. The event timing, $t_{\mathrm{max}}$, was defined by the timestamp of the maximum amplitude observed across the array. This value was used to fix the integration limits to $[t_{\mathrm{max}} - 20, t_{\mathrm{max}} + 200]$\,ns, ensuring complete charge collection for LYSO pulses while limiting contributions from noise.

\subsection{Crystal calibrations} \label{sec:calibrations}

The calibration of the full detector system was performed in three successive steps:
\begin{itemize}
    \item Set the PMT high voltages using the LYSO intrinsic radioactivity and a $^{60}$Co source for the NaI(Tl) detectors. 
    \item Refine the inter-calibration of the LYSO and NaI(Tl) detectors using the \SI{511}{keV} positron–electron annihilation peak.
    \item Determine final LYSO calibration coefficients by performing an in-situ numerical minimization of the energy resolution of the full array.
\end{itemize}
Each step of the calibration procedure is described in detail in the paragraphs 
below.\\

After installation, the PMT gains for the LYSO array were equalized using the crystal's intrinsic natural radioactivity, adjusting the high voltage such that the self-activity peak at $\sim$\SI{0.60}{MeV} aligned with a signal amplitude of \SI{10}{mV},  corresponding to a peak anode current of \SI{0.2}{mA}. This procedure resulted in a high voltage of \SI{1.07}{kV} for the \textit{Pent}, and high voltages between  1.1 and \SI{1.2}{kV} for the \textit{Hex} crystals. With these values, a peak anode current of \SI{33}{mA} is expected at \SI{100}{MeV}, well within the PMT's linear range. For the surrounding NaI(Tl) crystals, the initial gain was established using a $^{60}$Co source by setting the \SI{1.33}{MeV} photopeak to \SI{50}{mV}.

In the offline analysis, the inter-calibration of the NaI(Tl) detectors and LYSO crystals was refined using the \SI{511}{keV} positron-electron annihilation peak. This peak can be observed in the low-energy part of each crystals energy spectrum when no hodoscope position and NaI(Tl) energy cut is required, and its position was determined by a Gaussian fit. A rate-dependent gain shift was observed in the \SI{1.5}{"} PMT coupled to the central $\textit{Pent}$ crystal, scaling with the beam rate due to changes in the average anode current, as shown in appendix \ref{app:pmt_gain}. This response was successfully corrected in offline analysis by calibrating to the \SI{511}{keV} photopeak at each beam energy.

The final inter-calibration of the LYSO crystals was performed using an in-situ numerical minimization of the energy resolution. For each event $k$, the total reconstructed energy $E_{\text{rec}}^{(k)}$ was defined as the linear combination of the energy deposited in each individual LYSO channel $i$, weighted by a calibration coefficient $c_i$, plus the energy deposited in the NaI(Tl) and upstream detectors:
\begin{equation}
    E_{\text{rec}}^{(k)} = \sum_{i} c_i E_{i}^{(k)} + E_\text{rest}^{(k)}
\end{equation}
where the coefficients $c_i$ were treated as free parameters. $E_\text{rest}^{(k)}$ was calculated as the sum of energies deposited in the NaI(Tl)-detectors calibrated with the \SI{511}{keV} photopeak, and the energy deposited in the upstream detectors calibrated by scaling the observed energy spectrum to the expected value from \Gfour simulations.  The objective function for the minimization was defined as the fractional energy resolution, $R$, calculated as the ratio of the standard deviation to the mean of the reconstructed energy distribution $R(c) = {\sigma(E_{\mathrm{rec}})}/{\mu(E_{\mathrm{rec}})}$. To mitigate the bias introduced by the low-energy tail associated with lateral leakage, the resolution $R$ was evaluated only on the subset of events $\Omega'$ exceeding the 50th percentile, $P_{50}$, of the current energy distribution:

\begin{equation}
    \Omega' = \{ E_{\text{rec}}^{(k)} \mid E_{\text{rec}}^{(k)} > P_{50}(E_{\mathrm{rec}}) \}.
\end{equation}
The minimization procedure was initialized with unity gains ($c_i = 1$) and was performed at each beam energy separately. Testing the minimization procedure on Geant4 simulation data of the prototype detector geometry showed that the minimization procedure recovered the true relative weights $c_i$ to within \SI{0.25}{\percent}, while introducing a negligible bias in the resulting energy resolution of approximately \SI{0.01}{\percent} in absolute terms as shown in appendix \ref{app:calibration}.

\begin{figure}[h]
    \centering
    \includegraphics[width=0.6\textwidth]{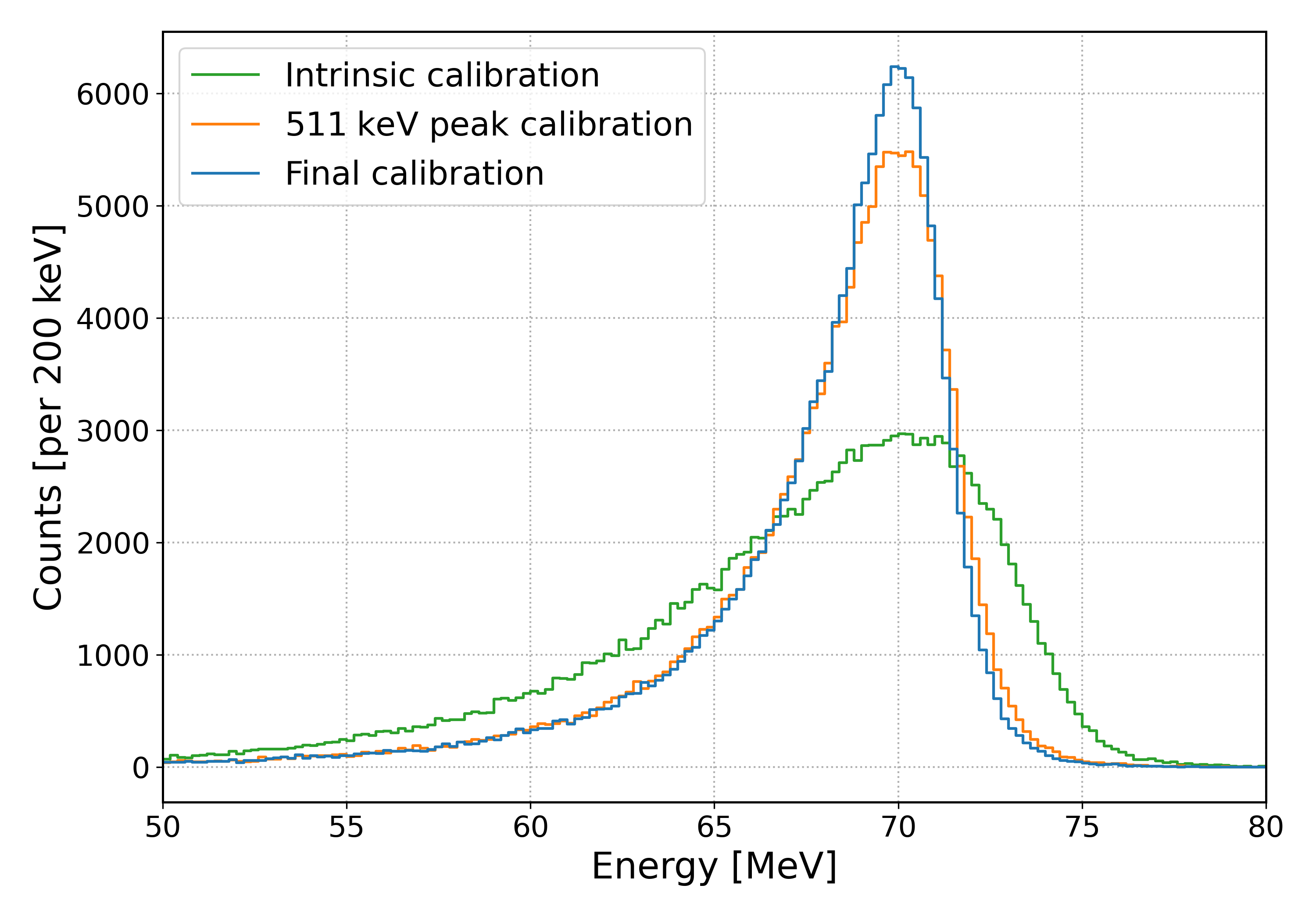}
    \caption{Comparison of the energy spectra from 70 MeV positrons obtained after adjusting the PMT high voltages using the LYSO intrinsic radioactivity peak (green), with refined calibration using the \SI{511}{keV} photopeak (orange), and with the final calibration after minimization of the standard deviation (blue). The intrinsic calibration spectrum shows the summed pulse integrals with a common global scaling factor applied to match the nominal beam energy scale for visualization purposes.}
    \label{fig:Calibration_comparison}
\end{figure}

\begin{figure}[h]
    \centering
    \includegraphics[width=0.8\textwidth]{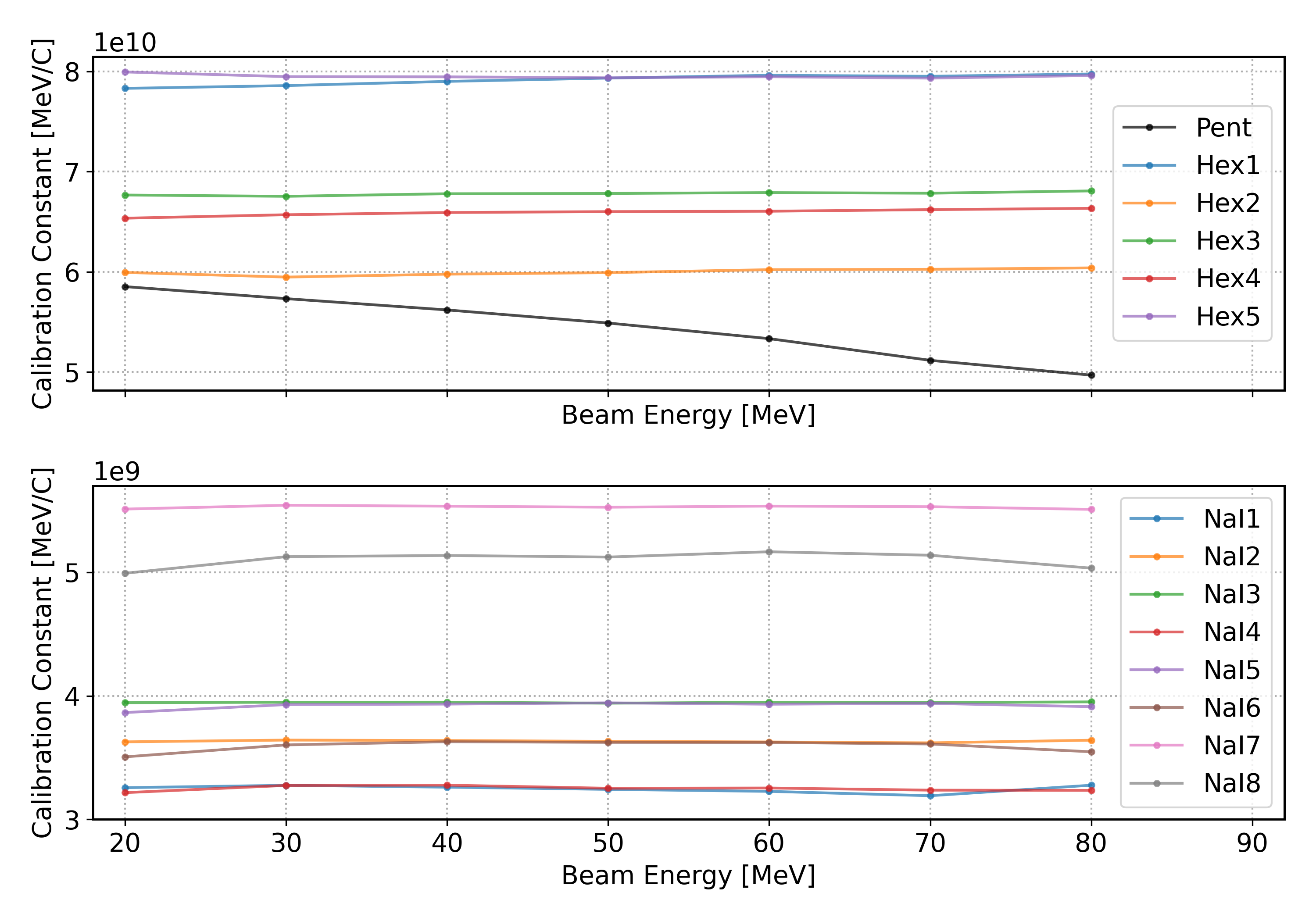}
    \caption{Final calibration constants of all LYSO and NaI detectors, obtained after the full calibration procedure. While the \textit{Hex} and NaI(Tl) calibration constants remain stable over the full energy range, the \textit{Pent} calibration shows a strong energy dependence. This was caused by the higher beam rates at higher energies and a rate dependent gain shift of the PMT, as shown in Appendix \ref{app:pmt_gain}.}
    \label{fig:Final_calibration_constants}
\end{figure}

Multiplying the initial calibration constants, determined from the \SI{511}{keV} positron annihilation peak, by the relative LYSO weights $c_i$ obtained from the minimization procedure yields the final calibration constants shown in figure \ref{fig:Final_calibration_constants}, which directly convert the measured pulse integrals into reconstructed energy. Using these final calibration constants, we find that the reconstructed peak positions are not exactly aligned with the expected positions from the Geant4 simulations at the nominal beam energies. The offsets have a standard deviation of \SI{0.55}{MeV} and may in part be related to the operation of the beamline below its nominal energy range of 100 to \SI{500}{MeV} \cite{Cline:2022pim1}, where the magnet response could be affected by hysteresis or limitations in the stability of the power supplies. We included this uncertainty on the reconstructed energy as horizontal error bars in the energy resolution parametrization in figure \ref{fig:energy_resolution_vs_energy}. A comparison of the energy spectra obtained after the different calibration steps is shown in figure \ref{fig:Calibration_comparison}.\\
Because the calibration was performed at each energy separately, the rate-dependent PMT gain, as well as gain drifts due to temperature variations on longer time scales than the $\sim$\SI{1}{h} runs, are calibrated out, while gain instabilities on shorter time scales still contribute to the measured energy resolution. However, splitting the datasets into smaller subsets of 100000 events and performing the calibration on each subset individually did not reveal any instabilities on shorter timescales.

\subsection{Data quality cuts}\label{sec:cuts}
Three basic selection cuts were applied to mitigate beam backgrounds and to ensure a single positron event in each digitization window. A timing cut relative to the cyclotron RF was used to reject background muons and pions by requiring the signal to lie within the main positron relative phase peak, with boundaries defined at \SI{10}{\percent} of the peak height on both sides. Pileup events were rejected by identifying double pulses in the $\mathrm{T0}$ detector or additional small signals in the Veto detector that did not trigger a veto signal. In addition, events with anomalously large pulses in the $\mathrm{T0}$ detector or hodoscope were removed from the analysis by removing events above the 90th percentile. Finally, a hodoscope position cut was applied to restrict the impact position in a circle of \SI{7}{mm} radius around the center of the $\textit{Pent}$ crystal, where lateral shower leakage is minimal. The hodoscope alignment relative to the $\textit{Pent}$ crystal was checked for each run by first binning the events according to their impact position measured by the hodoscope. Then, the mean energy deposited in the $\textit{Pent}$ crystal was calculated for each hodoscope pixel. The resulting two-dimensional energy distribution was fitted with a two-dimensional Gaussian to determine the position corresponding to the $\textit{Pent}$ center.

%% file: Sections/results.tex
\section{Results}
\label{sec:results}

\subsection{Energy resolution}

\begin{figure}[h]
    \centering
    \includegraphics[width=0.8\textwidth]{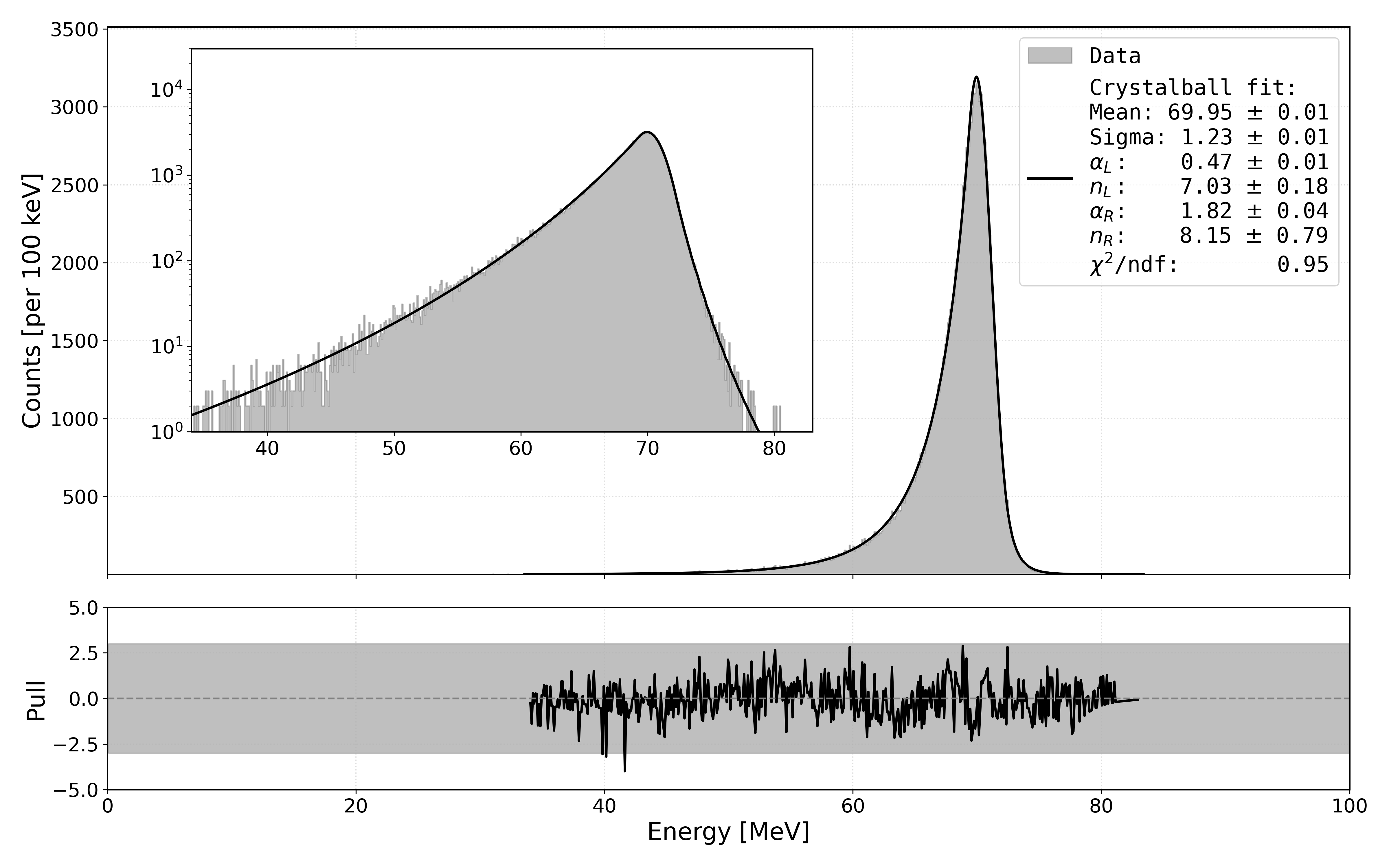}
    \caption{Energy distribution at \SI{70}{MeV}, fitted with a double-sided Crystal Ball function. From the standard deviation and mean of the Gaussian core, an energy resolution of $(1.75 \pm 0.02)$\,\si{\percent} was obtained. A small number of events exhibit anomalously high recorded energies. These events are due to the non-uniformity of the \textit{Pent} crystal: stochastic fluctuations in shower development can result in showers with significant energy deposition near the crystal's photosensor face, where the relative light response is higher.}
    \label{fig:energy_resolution_70MeV}
\end{figure}

\begin{figure}[h]
    \centering
    \includegraphics[width=0.6\textwidth]{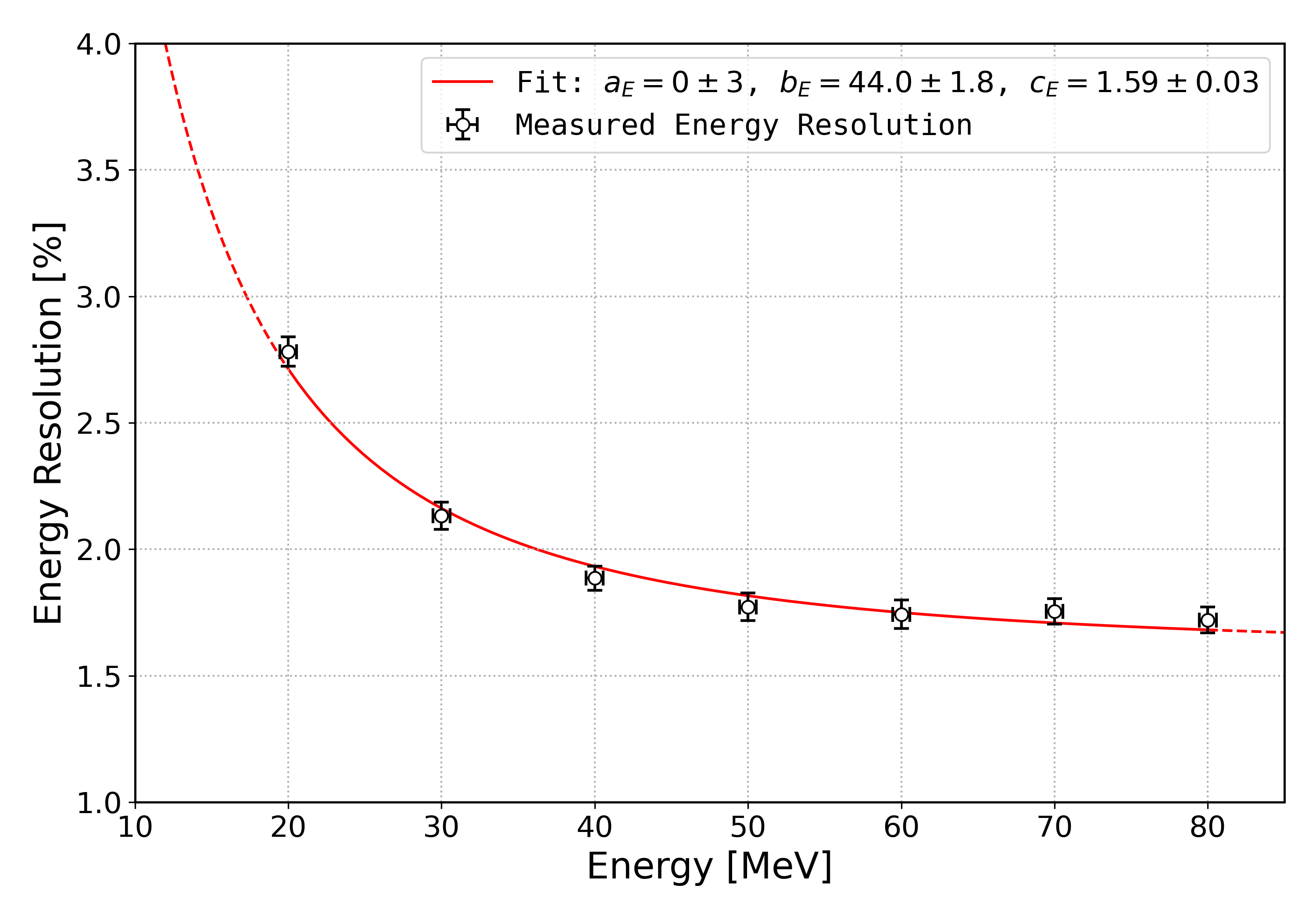}
    \caption{Energy resolution of the LYSO calorimeter after calibration, ranging in energy from \SI{20}{MeV} to \SI{80}{MeV}, with a fit to equation \eqref{eq:eres_e}. The vertical error bars here represent the spread in energy resolution observed when varying the calibration constants by \SI{1.5}{\percent}, which is their average variation along the energies. This serves as an estimate of the remaining uncertainty associated with miscalibrations or gain variations. The statistical uncertainties calculated from the crystal ball fit parameters are much smaller and decrease from \SI{0.04}{\percent} at \SI{20}{MeV} to \SI{0.015}{\percent} at \SI{80}{MeV}. The horizontal error bars represent the standard deviation of the differences between the reconstructed peak energies and their expected values. }
    \label{fig:energy_resolution_vs_energy}
\end{figure}

The energy resolution was determined at beam energies ranging from 20 to \SI{80}{MeV} in steps of \SI{10}{MeV}. Lower beam energies were not included because the positron beam rate decreased substantially with decreasing energy making it impractical to collect sufficient statistics, while the upper limit of \SI{80}{MeV} was chosen to extend just beyond the PIONEER signal region. In addition to the event selection cuts described above, a loose NaI requirement was imposed, limiting the energy deposited in the NaI detectors to less than \SI{20}{\percent} of the total event energy to ensure the energy resolution is dominated by the LYSO array. For each energy $E$, the summed energy deposition spectrum of the LYSO array was fitted with a double-sided Crystal Ball function \cite{Oreglia:CrystalBall} in the region $[0.5 \times E, 1.2 \times E]$, and the resolution was extracted as $\sigma_E / E$ from the fit. The energy dependence of the resolution was then described using the parametrization 
\begin{equation}
    \frac{\sigma_E}{E} (\%) = \frac{a_E}{\sqrt{E}} \oplus \frac{b_E}{E} \oplus c_E,
    \label{eq:eres_e}
\end{equation}
where the stochastic term $a_E$, noise term $b_E$, and constant term $c_E$, were determined from a fit to the measured data. The obtained fit parameters are $a_E = (0.0 \pm 3.2)\,\sqrt{\mathrm{MeV}}$, $b_E = (44.0 \pm 1.8)$ \si{MeV} and $c_E = 1.59 \pm 0.03 $. The measured resolution is below \SI{2}{\percent} for beam energies above \SI{40}{MeV}, satisfying the needs for a PIONEER calorimeter. At these energies, the constant term $c_E$ dominates. This term combines all contributions that produce absolute fluctuations that scale linearly with energy, such as lateral shower leakage, PMT miscalibrations, voltage/ gain instabilities, crystal response non-uniformity, beam momentum spread or energy deposition in dead material. The energy resolution reported here characterizes the performance of the prototype under realistic experimental conditions and we therefore do not attempt to subtract these contributions. A higher constant term compared to our 2023 measurement was expected from \Gfour simulations, mostly due to the limited lateral size of the array and the longitudinal non-uniformity. An improvement in comparison to our previous measurement was observed at low energies, reflected in the smaller noise term $b_E$, which we attribute to the use of new PMTs. The stochastic term is consistent with zero, which we interpret as the consequence of the limited sensitivity to this contribution over the measured 20 to \SI{80}{MeV} range. We therefore recommend that the parametrization should not be used to extrapolate to substantially lower energies, where stochastic contributions become more significant. An example energy distribution at \SI{70}{MeV} and the energy dependence of the energy resolution are shown in figures \ref{fig:energy_resolution_70MeV} and \ref{fig:energy_resolution_vs_energy} respectively.

The $R_{\pi}$ ratio $\Gamma(\pi\to e\nu)/\Gamma(\pi\to\mu\nu)$ is determined by associating events with either the $\pi \rightarrow e $ branch with its \SI{69.8}{MeV} monoenergetic positron or the $\pi\to\mu\nu$ branch, where the muon subsequently decays to a positron in the range from 0 to \SI{52.8}{MeV}. An adjustable energy cut, nominally at \SI{56}{MeV}, distinguishes the processes. A low-energy tail from the incomplete containment of the \SI{69.8}{MeV} energy in the $\pi\to e\nu$ branch extends below the cut and overlaps the dominant Michel decay spectrum. The tail fraction defined as  $c_\text{tail} = \frac{N[0 < E < 56\mathrm{MeV}]}{N[0 < E < 100\mathrm{MeV}]}$ for the \SI{70}{MeV} data shown in figure \ref{fig:energy_resolution_70MeV} is $c_\text{tail} = (2.2\pm0.1)$\si{\percent}; the uncertainty is purely statistical, and the fraction agrees with the integrated line shape from the fit shown. \Gfour simulations for the prototype array surrounded by the NaI detectors agree well with the measured results and clearly show that the tail fraction is dominated by lateral shower leakage, as expected. PIONEER requires a tail fraction for the full calorimeter to lie below \SI{1}{\percent} and be determined to a precision of \SI{1}{\percent}. Simulations of the full calorimeter geometry find a tail fraction well below that target. It will be measured directly using a positron beam and monitored continuously during pion stopping runs by suppressing Michel events using topology cuts in the active stopping target.

\subsection{Time resolution}

\begin{figure}[h]
    \centering
    \includegraphics[width=0.6\textwidth]{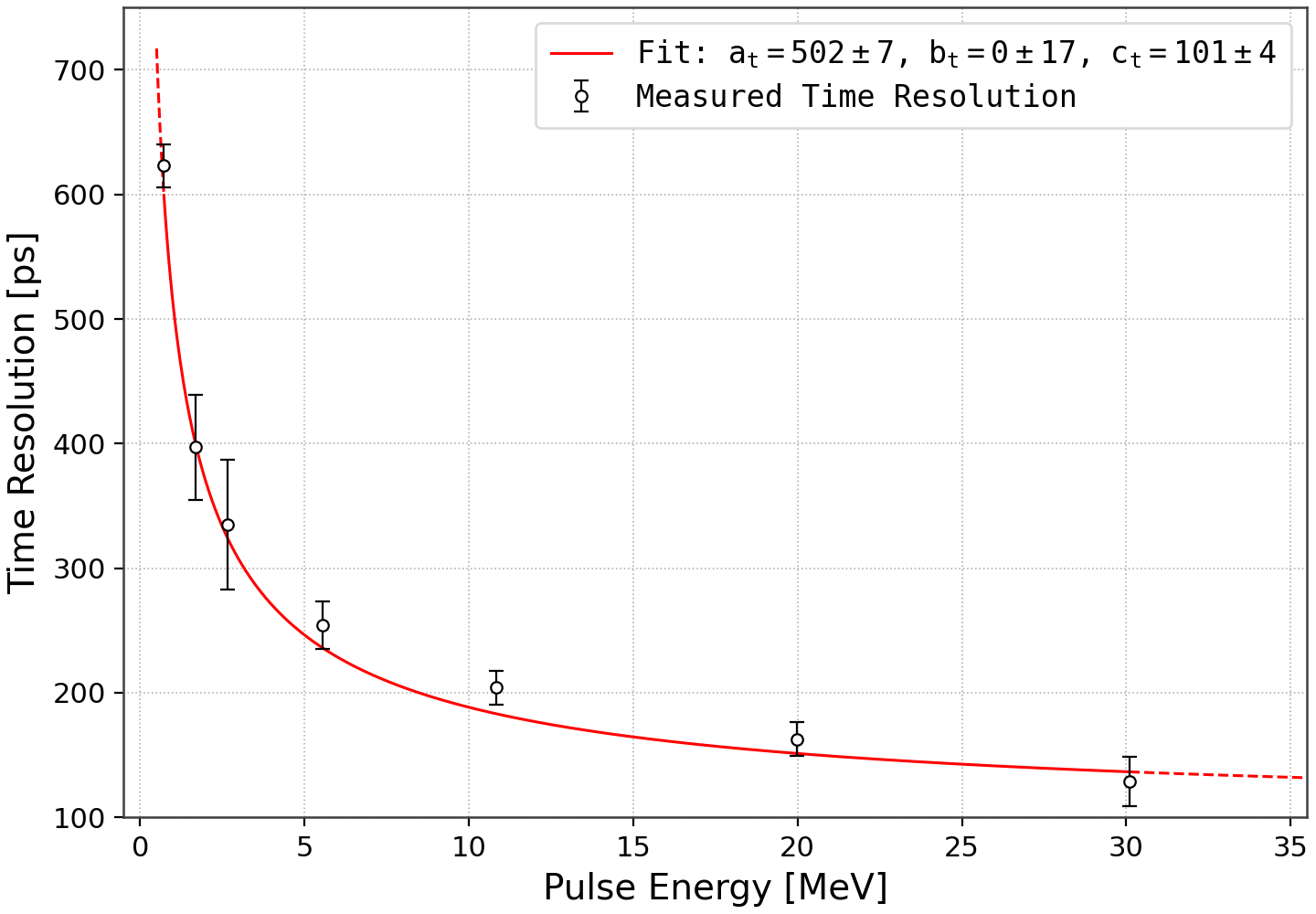}
    \caption{Time resolution of LYSO hits ranging in energy from 0.5 to \SI{30}{MeV} after removing the reference hit resolution via quadrature, with a fit to equation \eqref{eq:tres_e}.}
    \label{fig:time_resolution}
\end{figure}

A time resolution analysis was performed on the data collected at a beam energy of \SI{70}{MeV} with the RF and pileup cuts described in section \ref{sec:cuts} applied. The time resolution was determined for individual LYSO waveforms rather than for the combined signal of the array. Due to the speed of shower development, the time difference between hits in adjacent crystals has a small time spread compared to the calorimeter time resolution, which enabled the use of a reference pulse in the \textit{Pent} to extract the time resolution of lower energy hits in the neighboring \textit{Hex} crystals. To establish a precise reference time $(t_\text{ref})$, an additional event-level cut required an energy deposit of at least \SI{30}{MeV} in the \textit{Pent}. Consequently, the energy deposited in the second crystal could reach at most \SI{40}{MeV}. However, \SI{30}{MeV} was the highest energy at which sufficient statistics were available for a reliable determination of the time resolution. A template waveform was constructed for each channel by averaging many normalized, time-aligned pulses. Individual hit waveforms were then fit to this template with the amplitude and a time offset as free parameters, minimizing the squared residuals over the pulse region. The fitted time offsets relative to the template reference yield the hit times $(t_\text{ref})$ in the \textit{Pent} and $t_\text{signal}$ in the \textit{Hex} crystals. The time differences, defined as $\Delta t= t_\text{ref} - t_\text{signal},$ were histogrammed for various signal energies to capture the energy dependence of the time resolution. The intrinsic time resolution of the LYSO array was determined by extracting the standard deviation of the $\Delta t$ distributions across the 0.5 to \SI{30}{MeV} energy range, and subsequently subtracting the timing contribution of the reference pulse, determined through a three-detector method, in quadrature. The energy dependence of the time resolution was then fit to the equation \begin{equation}
    \Delta t = \frac{a_{t}}{\sqrt{E}} \oplus \frac{b_t}{E}  \oplus c_{t}.
    \label{eq:tres_e}
\end{equation}
The obtained fit parameters are $a_{t} = (502 \pm7)$ $\sqrt{\text{MeV}}$ $\cdot$ ps, $b_{t} = (0 \pm17)$ MeV $\cdot$ ps,  and $c_t = (101 \pm4)$ ps. Here, the first term encapsulates stochastic contributions, the second term captures noise, and the third term captures contributions from clock jitter and synchronization. The energy dependence of time resolution is shown in figure \ref{fig:time_resolution}.

\subsection{Position resolution}

\begin{figure}[htbp]
    \centering
    \includegraphics[width=0.8\textwidth]{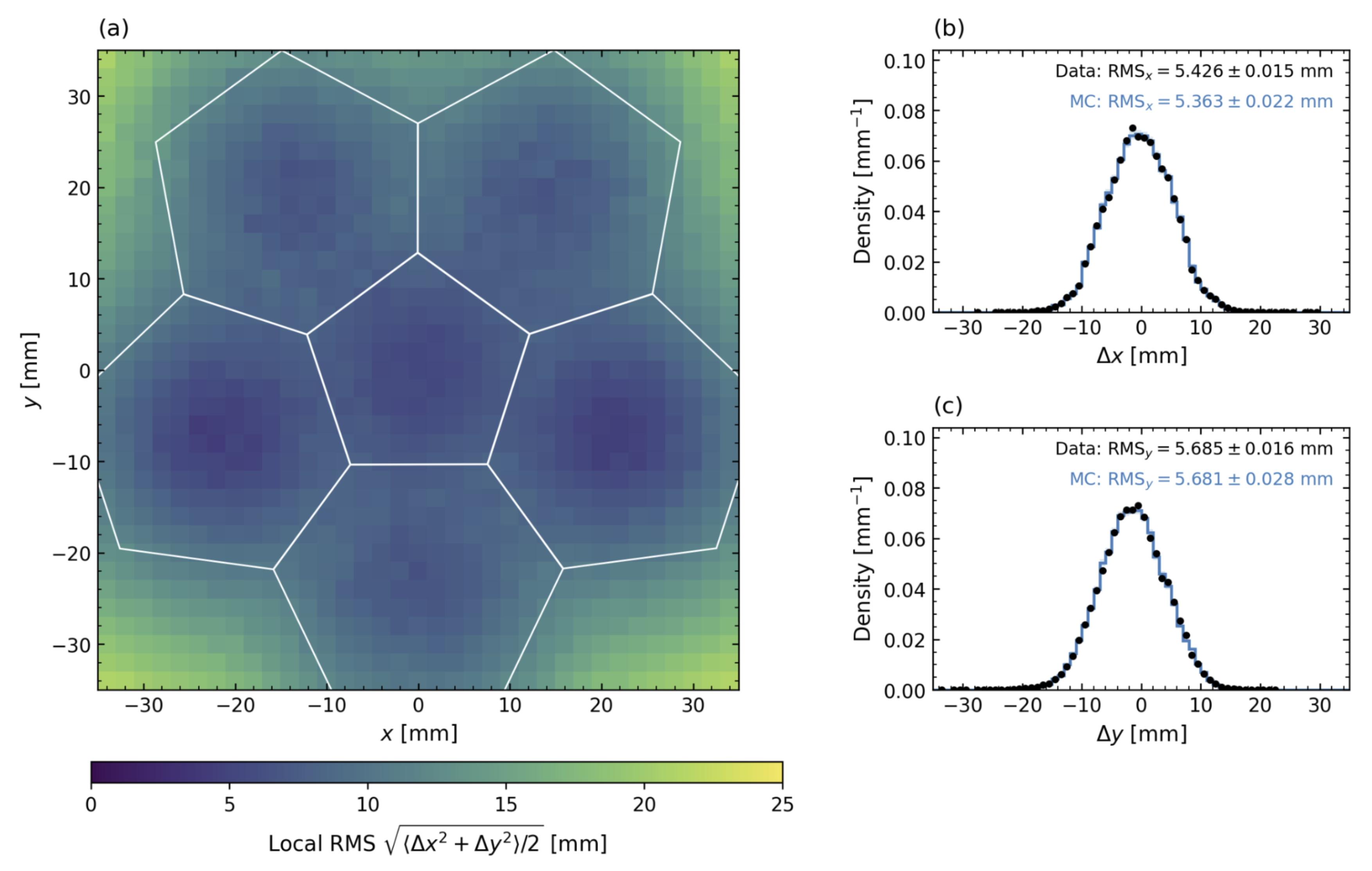}
    \caption{Panel (a) shows the local RMS of the position reconstructed by the logarithmic energy-weighted center-of-mass algorithm compared to the Monte Carlo truth position for positrons incident on the entire front face of the prototype. Position resolution degrades near the edges of the LYSO array, where the shower is not fully contained. Panel (b) and (c) compare the reconstructed position in x and y to Monte Carlo truth for simulation and to the position recorded by the hodoscope for data.}
    \label{fig:position_resolution}
\end{figure}

The silicon strip detector, positioned upstream of the LYSO array, provided a precise measurement of the (x,y) impact position of the incoming positrons. This was used as a reference for evaluating the position reconstruction performance of the LYSO array using a logarithmic energy-weighted center-of-mass formula as described in \cite{AWES1992130}. The logarithmic weighting parameter $w_0$ was scanned in the range of 1.0 to 5.0 using a \Gfour simulation at \SI{70}{MeV}, from which an optimal value of $w_0 = 2.6$ was found for the prototype geometry. In the region covered by the hodoscope, the RMS of the residuals between the reconstructed and reference positions are \SI{5.43}{mm} and \SI{5.69}{mm} in x and y, respectively, for data, as shown in figure \ref{fig:position_resolution} on the right.

To distinguish the local reconstruction bias from the residual spread, the events were further binned according to their hodoscope position. In each bin, the local bias was defined as the mean reconstructed-position residual, while the corresponding spread was defined as the standard deviation about this mean. After removing the local biases, the resulting standard deviations in the hodoscope coverage region are \SI{5.06}{mm} and \SI{5.20}{mm} in x and y, respectively. The variation of the local bias accounts for approximately \SI{15}{\percent} of the total variance in each direction.

The same analysis was performed using \Gfour simulation, allowing the position reconstruction performance in the hodoscope coverage region to be compared between data and simulation. The simulation was additionally used to evaluate the reconstruction performance for 70 MeV positrons incident on the full front face of the crystal array, beyond the hodoscope coverage region as shown in figure \ref{fig:position_resolution} on the left.
The analysis demonstrates that, at this energy, the highly asymmetric development of the positron showers limits the positional information gained from energy sharing between neighboring crystals. The logarithmic weighting algorithm therefore provides only a small improvement over assigning the event to the center of the crystal with the largest energy deposition.

%% file: Sections/conclusion.tex
\section{Conclusion}
\label{sec:conclusion}

Our studies of an array of six large, tapered LYSO crystals found:
\begin{itemize}
    \item The longitudinal response uniformity of each of the tapered crystals was found to be better than \SI{7}{\percent}, which is sufficient for use in the PIONEER calorimeter.
    \item The energy resolution of an array composed of these LYSO crystals surrounded by NaI tail catchers was mapped out from 20 to \SI{80}{MeV}, as shown in figure \ref{fig:energy_resolution_vs_energy}. An energy resolution of \SI{1.75}{\percent} was measured at \SI{70}{MeV}, exceeding the PIONEER specification despite degradation due to lateral leakage from the array.
    \item The time resolution of the pulses in the array was mapped out from 0.5 to \SI{30}{MeV}, as shown in figure \ref{fig:time_resolution}. A time resolution below \SI{140}{ps} was found for \SI{30}{MeV} pulses.
    \item The position resolution was measured to be \SI{5.06}{mm} in x and \SI{5.20}{mm} in y after removal of local biases in the central region of the array covered by a silicon strip detector providing precise position information.
\end{itemize}

The above results have shown that large LYSO crystals in the shapes and sizes currently needed for the PIONEER calorimeter meet the performance requirements of the PIONEER experiment. Future work will focus on further optimizing the crystal surface treatment and wrapping procedure to improve crystal response uniformity. To this end, crystals with different surface roughening schemes will be subjected to source scans. The results will be used to inform Geant4 optical simulations, which can be used to extrapolate the results to other crystal geometries. Additionally, the PMT voltage dividers will be further optimized to deliver improved pulse shapes for high time resolution, while maintaining linear response to pulses ranging in energy from \SI{0.1}{MeV} to \SI{100}{MeV}.

%% file: Sections/app-pmt.tex
\usetikzlibrary{positioning, shapes.geometric, arrows.meta, calc}
\definecolor{naiGray}{RGB}{80,80,80}
\definecolor{discOrange}{RGB}{210,120,0}
\definecolor{na22Green}{RGB}{80,160,50}
\definecolor{pentRed}{RGB}{180,30,30}
\definecolor{lysoBlue}{RGB}{20,60,140}
\definecolor{hexBlue}{RGB}{20,60,140}
\definecolor{ndCyan}{RGB}{0,160,200}
\definecolor{ledCyan}{RGB}{0,180,220}
\definecolor{wdNavy}{RGB}{15,50,120}
\definecolor{trigBox}{RGB}{230,240,255}

\pagebreak

\section{Verification of the PMT response} \label{app:pmt_gain}

\begin{figure}[htbp]
    \centering
    
    \begin{subfigure}{0.48\textwidth}
        \centering
        \resizebox{\textwidth}{!}{%
        \begin{tikzpicture}[
            font=\sffamily\bfseries,
            every node/.style={align=center},
            line width=1.2pt
        ]
        
        %
        %
        %
        
        \node[draw=naiGray, fill=naiGray, text=white,
            minimum width=7.5cm, minimum height=1.4cm, inner sep=6pt
        ] (NaI) at (3.0, 4.2) {NaI};
        
        \node[draw=discOrange, fill=white, text=discOrange,
            minimum width=1.8cm, minimum height=1.4cm, line width=2pt, inner sep=6pt
        ] (Disc) at (10.5, 4.2) {Disc.};
        
        \node[draw=na22Green, fill=na22Green, text=white,
            minimum width=1.6cm, minimum height=0.7cm, inner sep=4pt
        ] (Na22) at (3.0, 2.85) {Na22};
        
        \node[draw=pentRed, fill=white, text=pentRed,
            minimum width=2.6cm, minimum height=1.8cm, line width=2pt, inner sep=6pt
        ] (PENT) at (-0.9, 1.5) {H3178-51};
        
        \node[draw=lysoBlue, fill=lysoBlue, text=white,
            minimum width=5.2cm, minimum height=1.2cm, inner sep=6pt
        ] (LYSO) at (3.0, 1.5) {LYSO};
        
        \node[draw=hexBlue, fill=white, text=hexBlue,
            minimum width=2.6cm, minimum height=1.8cm, line width=2pt, inner sep=6pt
        ] (HEX) at (6.9, 1.5) {H13719};
        
        \node[draw=ndCyan, fill=ndCyan, text=white,
            minimum width=4.2cm, minimum height=0.7cm, inner sep=4pt
        ] (ND) at (3.0, 0.45) {ND Filter Wheel};
        
        \node[draw=ledCyan, fill=ledCyan, text=white,
            minimum width=2.2cm, minimum height=1.4cm, inner sep=6pt
        ] (LED) at (3.0, -0.7) {Blue\\LED};
        
        \node[draw=wdNavy, fill=wdNavy, text=white,
            minimum width=1.8cm, minimum height=4.0cm, inner sep=6pt
        ] (WD) at (10.5, 0.75) {DAQ};
        
        
        \draw[-] (NaI.east) -- (Disc.west);
        
        \draw[-] (Disc.south) -- (WD.north);
        
        \draw[-] (HEX.east) -- (WD.west |- HEX.center);
        
        \draw[-]
            (PENT.south)        
            -- (-0.9, -2.0)     
            -- (10.5, -2.0)     
            -- (WD.south);      
        \end{tikzpicture}%
            }
        \caption{}
    \end{subfigure}
    \begin{subfigure}[b]{0.48\textwidth}
        \centering
        \includegraphics[width=\textwidth]{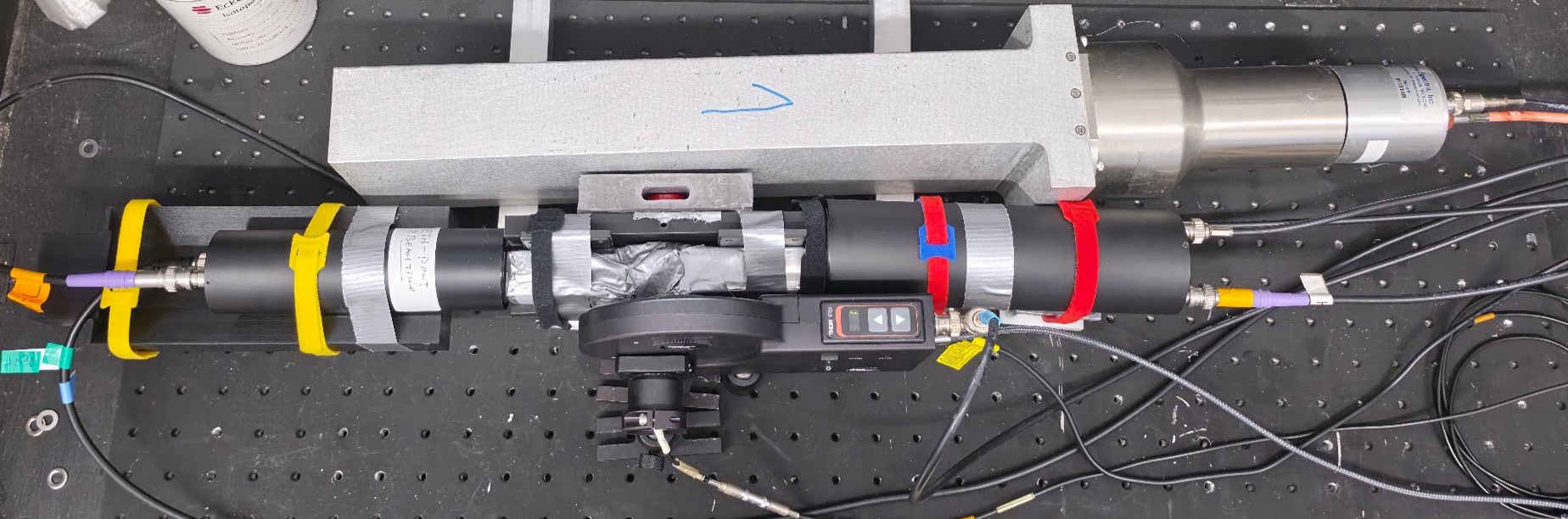}
        \caption{}
        \label{appfig:gain_setup_b}
    \end{subfigure}
    
    \caption{A diagram and photograph of the test stand. The lead-collimated $^{22}$Na source creates a coincidence between the LYSO signals and the NaI detector.}
    \label{appfig:gain_setup}
\end{figure}

\begin{figure}[htbp]
    \centering
    \includegraphics[width=\textwidth]{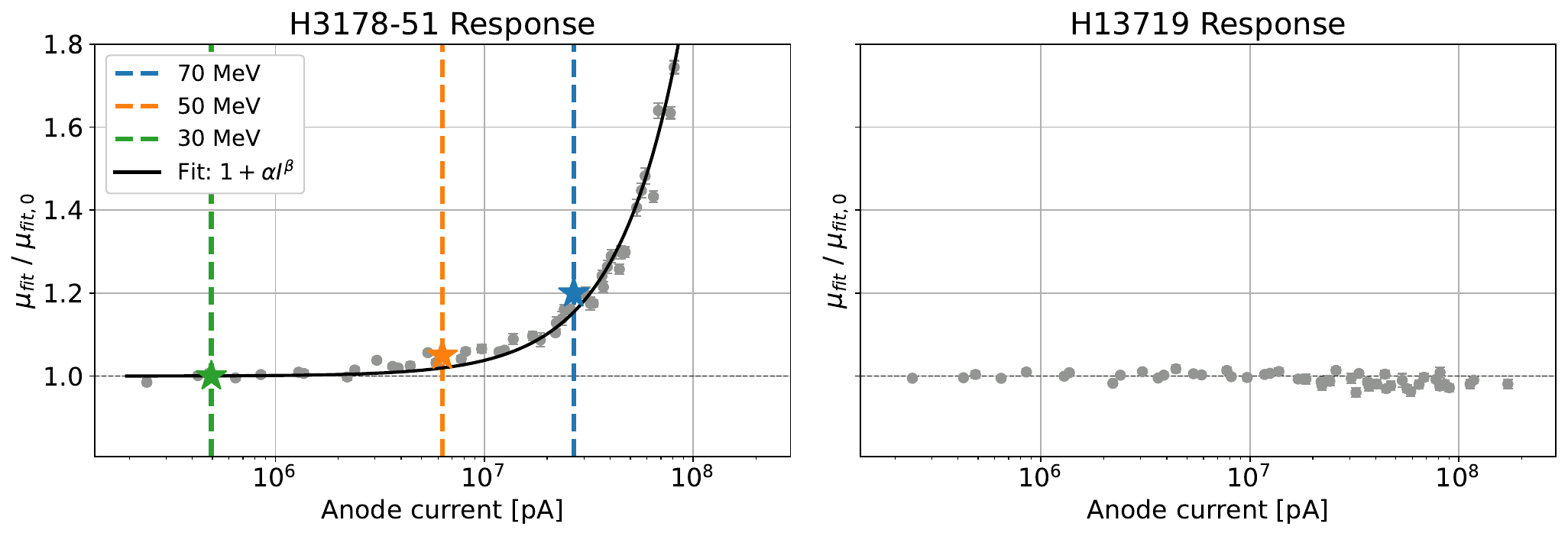}
    \caption{Fitted mean of the $^{22}$Na peak in the H3178-51 (\textit{Pent}, left) and H13719 (\textit{Hex}, right) PMTs vs. calculated time averaged anode current. The data shown here includes H3178-51 PMT high voltages from 1000 to \SI{1100}{V} and LED rates from \SI{100}{Hz} to\SI{36}{kHz}. The H13719 PMT high voltage was set to \si{1400}{V} for all tests. }
    \label{appfig:response_curves}
\end{figure}

After the conclusion of the measurements at PSI, a standalone test stand was assembled to investigate the rate-dependence of the gain that was observed for the PMT attached to the \textit{Pent} crystal, as mentioned in section \ref{sec:calibrations}.
Both types of PMTs used during the measurements at PSI were tested, one \SI{1.5}{"} Hamamatsu H3178-51 PMT (\textit{Pent}) and one \SI{2}{"} H13719 PMT (\textit{HexA}), both in the full assembly with magnetic shielding and tapered voltage divider. The test stand setup consisted of the two PMTs looking at opposite ends of a rectilinear LYSO crystal, a NaI detector, a collimated $^{22}$Na source, a variable-rate blue LED, and a variable array of neutral density (ND) filters. A schematic of the setup can be seen in Figure \ref{appfig:gain_setup}.
A hole was made in the ESR wrapping of the crystal to allow light from the blue LED to enter, and the LED was pulsed at rates $\Gamma_{\text{LED}} \in \left[\SI{1}{Hz}, \SI{30000}{Hz} \right]$. The pulse-by-pulse intensity was varied by changing the length of the pulse and varying the ND filter between the LED and the crystal face.
The coincidence of the NaI and LYSO signals from back-to-back $^{22}$Na $\gamma$ emission provided a signal mostly free of LED background, and simple analysis cuts removed any coincidences of the $\gamma$ and LED signals. 
The LED background was quantified in terms of the average anode current:
\begin{equation}\label{appeq:current}
    \bar{I}_{\text{anode}} \equiv Q_{\text{LED}} \times \Gamma_{\text{LED}},
\end{equation}
where $Q_{\text{LED}}$ was the integrated charge from a single LED pulse. The amplitude of the LED pulse was tuned to be approximately equal to the highest energy LYSO pulses when the lowest attenuation ND filter was in place. At each $\bar{I}_{\text{anode}}$, the mean of the \SI{511}{keV} photopeak was extracted from a Gaussian fit.
The relative responses of the two PMTs, normalized by the LED-Off measurements, are shown in Figure \ref{appfig:response_curves} for anode currents between $10^5-10^8$\,pA.
No deviation from unity in either PMT is seen up to \textasciitilde\,$2 \times 10^{6}$\,pA. After this point, the response of the H3178-51 PMT increases following a power law: $1 + \alpha I^\beta$. Across the same range of currents, the response of the H13719 PMT remains consistent with unity.
The average anode currents were not measured directly during the PSI measurements. The average in-situ anode current was therefore calculated by taking the average pulse integral for a data run and multiplying by the trigger rates. The gain shift for the PSI data was calculated by fitting the \SI{511}{keV} photopeak in the low-energy spectrum.
The measured currents and gain shifts for three data runs are plotted in figure \ref{appfig:response_curves} alongside the LED data. The in-situ response matches well with the response expected from the LED gain measurements. This data indicates that the H3178-51 PMT was operated in the over-linearity regime, where the voltage distribution gets distorted towards earlier dynode stages by the large average anode current, leading to an effective gain increase, as described in section 5.1.3 of ref. \cite{hamamatsu_handbook}. We therefore conclude that the observed gain shift was due only to PMT effects and as such applied a calibration factor to the data in the main analysis to correct for it.

\begin{figure}[h]
    \centering
    \includegraphics[width=0.6\textwidth]{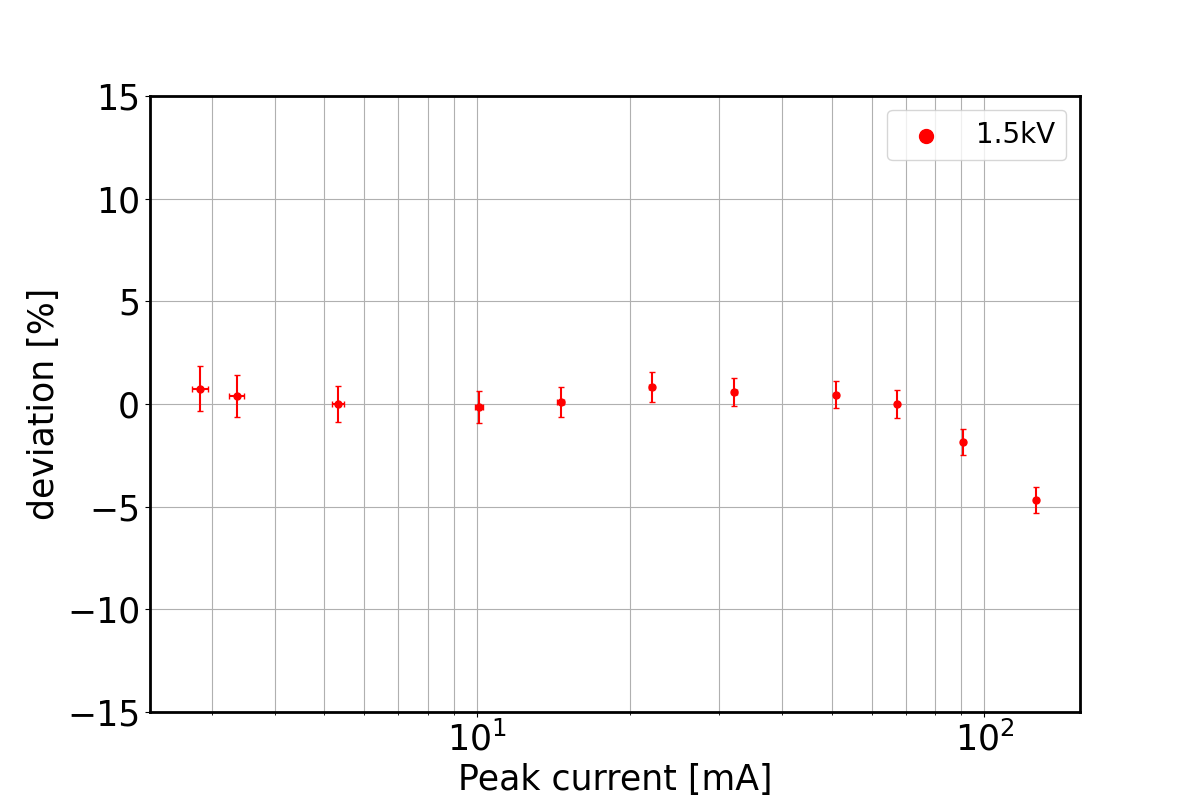}
    \caption{Single pulse linearity of the 2" H13719 PMT measured in a test setup before the PSI measurement. The response is linear ($<$ \SI{2}{\percent} deviation) up to a peak anode current of \SI{90}{mA}.}
    \label{fig:PMT_linearity}
\end{figure}

When subjected to high energy events, PMT signals can saturate because of space charge effects. This is independent from the rate dependent effect described above, and results in a non linear response where the effective gain decreases with increasing event energy. Before the beam time at PSI, the pulse linearity of the 2" H13719 PMT was investigated to confirm the performance of the custom made voltage divider circuit. To this end, the method described in section 4.3.3b of ref. \cite{hamamatsu_handbook} was used, where the response to two LED light pulses with a fixed intensity ratio of $\sim4:1$ at varying distances is compared. It was found that the deviation from linearity exceeds \SI{2}{\percent} for peak anode currents above \SI{90}{mA}, as shown in figure \ref{fig:PMT_linearity}. This is consistent with the \SI{100}{mA} stated in the data sheet provided by Hamamatsu, considering the uncertainty on the PMT gain, which was not measured independently.

%% file: Sections/app-calibration.tex
\section{Minimization calibration procedure} \label{app:calibration}

The minimization calibration procedure was tested in a Geant4 simulation of the detector including all LYSO and NaI(Tl) crystals and upstream detectors. Positron beams of the same energies (20 to \SI{80}{MeV} in \SI{10}{MeV} steps) and spot sizes as measured at PSI were simulated. Each crystal was assigned a relative calibration drawn from a Gaussian distribution centered at 1 with a width of 0.1, corresponding to a pessimistic 10\% relative calibration from the initial calibration on the \SI{511}{keV} positron annihilation peak. At each energy, the minimization converged to calibrations that were within \SI{0.25}{\percent} of the correct values. The relative calibrations before and after the minimization procedure are shown at each energy in Table \ref{tab:Calibration_simulation}. The final calibrations were nearly identical between any of 10 seeds tested.

\begin{table}[h]
    \centering
    \caption{Calibration constants obtained with the minimization procedure at all beam energies, starting from the same initial relative gains shown in the top row. Testing the procedure with varying initial gains yielded consistent results.}
    \label{tab:Calibration_simulation}
    \begin{tabular}{|c|c c c c c c|}
        \hline
        Constants & Pent & Hex1 & Hex2 & Hex3 & Hex4 & Hex5 \\
        \hline
        Initial & 1.0371 & 0.9869 & 0.9618 & 0.9849 & 0.9224 & 1.0099 \\
        20 MeV   & 0.9985 & 1.0009 & 1.0014 & 1.0003 & 1.0012 & 1.0002 \\
        30 MeV   & 0.9987 & 1.0007 & 1.0005 & 1.0007 & 1.0013 & 1.0006 \\
        40 MeV   & 0.9985 & 1.0014 & 1.0012 & 1.0002 & 1.0006 & 1.0014 \\
        50 MeV   & 0.9980 & 1.0013 & 1.0012 & 1.0020 & 1.0008 & 1.0013 \\
        60 MeV   & 0.9978 & 1.0018 & 1.0015 & 1.0014 & 1.0015 & 1.0015 \\
        70 MeV   & 0.9978 & 1.0018 & 1.0015 & 1.0010 & 1.0020 & 1.0016 \\
        80 MeV   & 0.9981 & 1.0007 & 1.0018 & 1.0013 & 1.0021 & 1.0011 \\
        \hline
    \end{tabular}
\end{table}
    
\begin{figure}[h]
        \centering
        \includegraphics[width=0.8\textwidth]{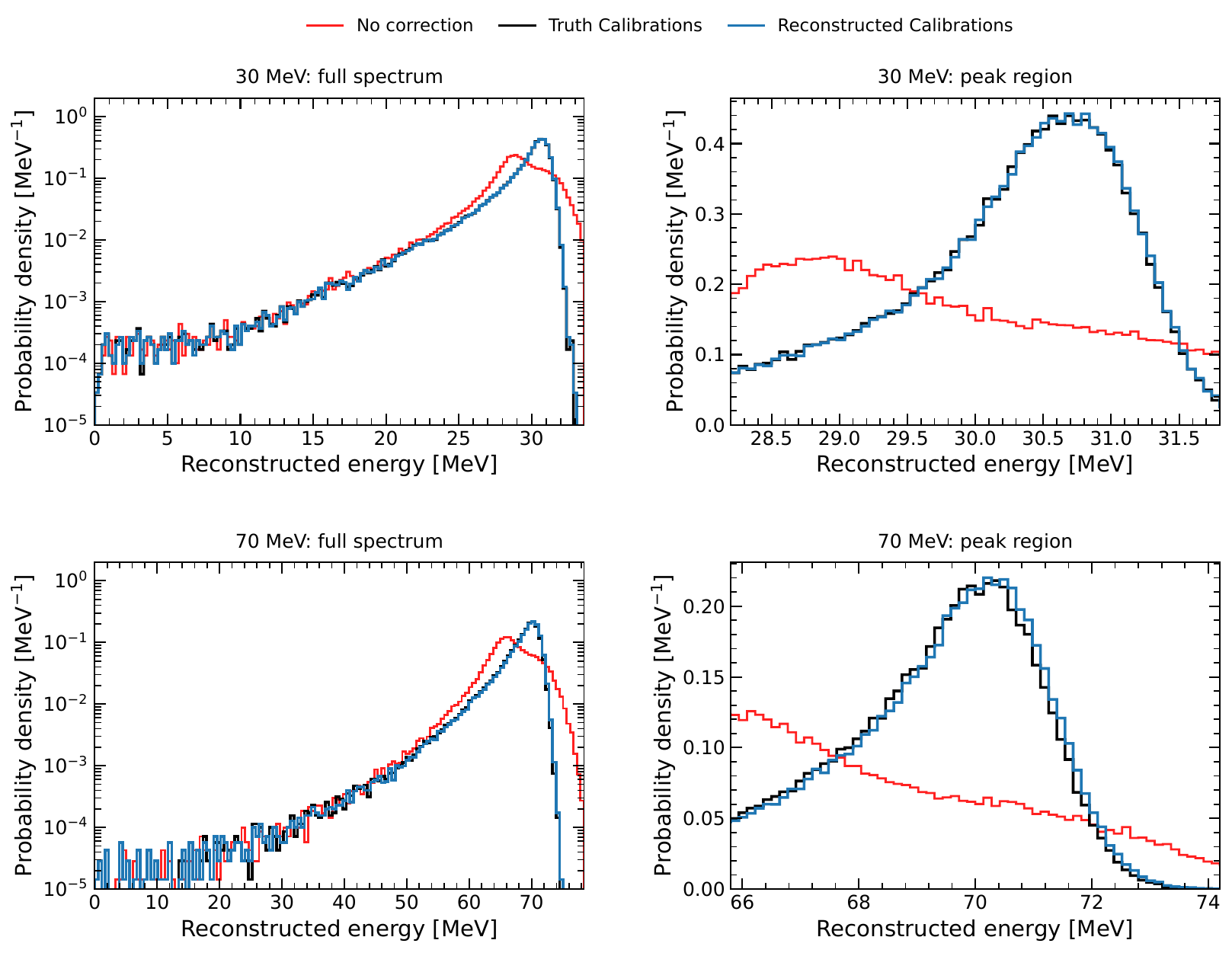}
        \caption{Comparison of the energy spectra at \SI{30}{MeV} and \SI{70}{MeV} obtained with: the relative calibrations drawn from a Gaussian distribution centered at 1 with a width of 0.1 (red), perfect calibrations using \Gfour truth information (black) and calibrations using the minimization procedure (blue).}
        \label{fig:minimization_procedure}
    \end{figure}

We then compared the energy spectra at each energy when the crystals were perfectly calibrated to when they were calibrated using the minimization procedure. The energy resolution was shifted by $~$\SI{0.01}{\percent} ($~$\SI{0.5}{\percent} relative) due to the procedure and we therefore conclude that its effect on the measured energy resolution is negligible. The comparison of the obtained energy spectra at beam energies of \SI{30}{MeV} and \SI{70}{MeV} is shown in figure \ref{fig:minimization_procedure}.

\begin{figure}[h]
        \centering
        \includegraphics[width=0.6\textwidth]{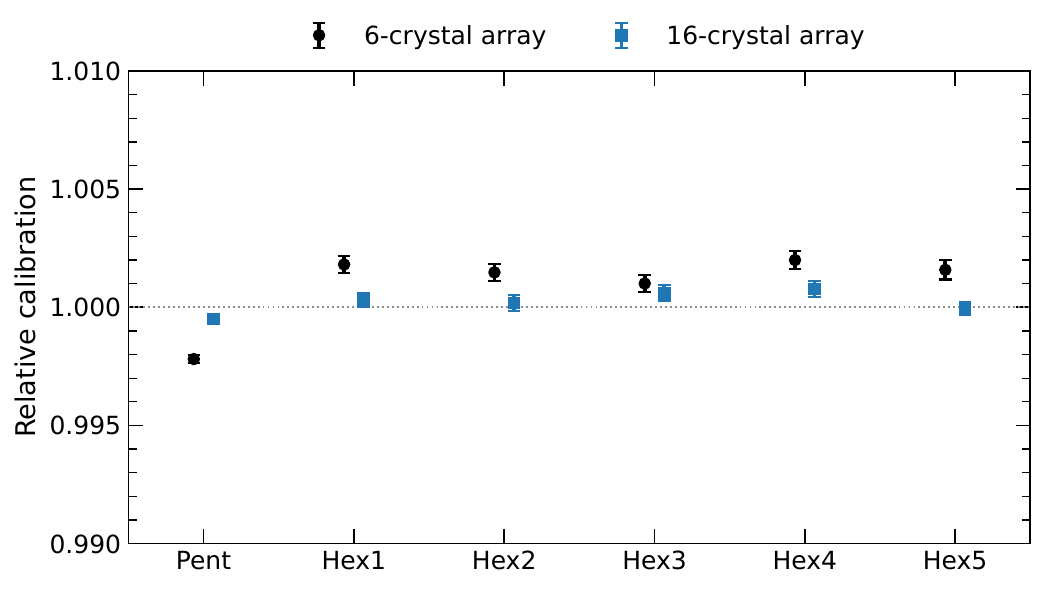}
        \caption{Comparison of the calibration bias observed in simulation of the 6-array, in comparison to the 16-array. In the larger array with better lateral containment, the bias decreased significantly.}
        \label{fig:calibration_bias}
    \end{figure}

Though the minimization introduces a negligible bias to the energy resolution, we see that the \textit{Pent} is systematically assigned a calibration 0.2 to \SI{0.4}{\percent} below the calibrations of the \textit{Hex} crystals in simulation. We observed that the effect is stronger if the minimization is performed on the full energy spectrum including the tail. We therefore believe that this miscalibration is due to lateral leakage from the array which the minimization tries to correct for by slightly increasing the relative responses of crystals near the edge of the array (effectively decreasing the relative response of the central crystal). To support this hypothesis, the calibration procedure was repeated in a simulation of a 16-crystal LYSO array where an additional ring of LYSO crystals surrounded the 6-crystal prototype, and the NaI(Tl) tail catchers were moved outward to accommodate the larger array. In this setup with better containment, this systematic miscalibration decreased significantly, as shown in figure \ref{fig:calibration_bias}.

